\documentclass[a4paper,11pt]{article}
\pdfoutput=1
\usepackage{jheppub}

\usepackage{amssymb,amsmath,amsfonts}
\usepackage[normalem]{ulem}
\usepackage[utf8]{inputenc}
\usepackage{slashed}
\usepackage{graphicx}
\usepackage{tabularx}
\usepackage{here}
\usepackage{color}
\usepackage{csquotes} 
\usepackage{comment}
\usepackage{mathrsfs}
\usepackage{mathcomp}
\usepackage{float}
\usepackage{ascmac}
\usepackage{multirow}
\usepackage{longtable}
\usepackage{bm}
\usepackage{ulem}
\usepackage{tikz}
\usetikzlibrary{positioning, arrows.meta, calc}
\usepackage{booktabs}
\usepackage{array}
\usepackage{caption}
\usepackage[italicdiff]{physics}

\newcounter{c1}
\newcounter{c}
\newcounter{c2}
\newcounter{c3}
\newcounter{c4}
\newcommand{\meV}{\mbox{meV}}
\newcommand{\eV}{\mbox{eV}}

\newcommand{\GeV}{\mbox{GeV}}

\makeatletter
\newcommand*\rel@kern[1]{\kern#1\dimexpr\macc@kerna}
\newcommand*\widebar[1]{%
  \begingroup
  \def\mathaccent##1##2{%
    \rel@kern{0.8}%
    \overline{\rel@kern{-0.8}\macc@nucleus\rel@kern{0.2}}%
    \rel@kern{-0.2}%
  }%
  \macc@depth\@ne
  \let\math@bgroup\@empty \let\math@egroup\macc@set@skewchar
  \mathsurround\z@ \frozen@everymath{\mathgroup\macc@group\relax}%
  \macc@set@skewchar\relax
  \let\mathaccentV\macc@nested@a
  \macc@nested@a\relax111{#1}%
  \endgroup
}
\makeatother

\numberwithin{equation}{section}

\preprint{
\begin{minipage}{5cm}
\small
\flushright
KYUSHU-HET-368\\
KUNS-3124
\end{minipage}} 

\title{
Uncovering Hidden Leptonic Correlations with Flow Matching and Autoencoders
}

\author{Haruto Kitagawa$^{1}$,}
\author{Satsuki Nishimura$^{2}$, and} 
\author{Hajime Otsuka$^{1,3}$} 
\affiliation{
$^1$Department of Physics, Kyushu University, \\ 744 Motooka, Nishi-ku, Fukuoka 819-0395, Japan}
\affiliation{
$^2$Department of Physics, Kyoto University, \\ Kitashirakawa-Oiwakecho, Sakyo-ku, Kyoto 606-8502, Japan
}
\affiliation{
$^3$Quantum and Spacetime Research Institute (QuaSR), Kyushu University, \\ 744 Motooka, Nishi-ku, Fukuoka 819-0395, Japan}
\emailAdd{kitagawa.haruto.691@s.kyushu-u.ac.jp}
\emailAdd{satsuki@gauge.scphys.kyoto-u.ac.jp}
\emailAdd{otsuka.hajime@phys.kyushu-u.ac.jp}

\abstract{
We perform a global search for values of the Yukawa matrices and Majorana masses in the Type-I seesaw mechanism. Using flow matching, which is a generative artificial intelligence (generative AI) method, we generate a broad set of solutions reproducing the experimentally measured values of the neutrino mass-squared differences and the mixing angles. Then, a machine learning method known as an autoencoder is applied to uncover non-trivial correlations among physical quantities in the lepton sector. Our analysis reveals new non-linear relations involving neutrino masses and CP phases. These findings may contribute to elucidating the origins of the mass hierarchies and mixing patterns among generation structure.
}

\makeatletter
\gdef\@fpheader{}
\makeatother

\begin{document}

\maketitle

\section{Introduction}
\label{sec:Intro}

The origin of fermion flavor remains one of the central unresolved problems in particle physics. 
Although the Standard Model successfully describes a wide range of phenomena, it does not explain the key factors underlying the observed hierarchies of fermion masses and mixing parameters. 
The lepton sector is particularly intriguing: the neutrino mass scale, neutrino mass ordering, the Majorana nature of neutrinos, and leptonic CP violation are not fully determined. 
Future oscillation, cosmological, beta-decay, and neutrinoless double-beta decay measurements are expected to provide increasingly stringent tests of models of lepton flavor \cite{Denton:2026wpy}.

\medskip

Bottom-up approaches have also been widely employed to connect low-energy neutrino data with the underlying model parameters. For example, Refs.~\cite{Hisano:1995cp,Hisano:1998fj,Casas:2001sr} reconstruct neutrino Yukawa couplings that reproduce the experimentally measured neutrino observables at low energies and subsequently investigate phenomenological consequences, such as charged-lepton flavor violation in supersymmetric models. Related approaches have also been applied to leptogenesis; for instance, Ref.~\cite{Falcone:2003af} discusses the baryon asymmetry by reconstructing high-energy model parameters from low-energy experimental inputs. 
At the level of the effective light-neutrino mass matrix, Ref.~\cite{Bertuzzo:2013ew} utilized oscillation data to construct probability distributions for the matrix elements and to analyze correlations among them in a data-driven approach.
In contrast, our aim is not to reconstruct a particular set of high-energy parameters, but to explore the global distribution of model parameters compatible with the low-energy neutrino data and to identify non-trivial correlations that emerge within the resulting parameter ensemble.
In this work, we consider the Type-I seesaw mechanism \cite{Minkowski:1977sc,Yanagida:1979as,Gell-Mann:1979vob,Mohapatra:1979ia} without imposing any flavor symmetry or specific texture on the Yukawa matrices. 
We identify the ranges of Yukawa couplings and Majorana masses that are consistent with the currently available neutrino data and study the relationships among the physical quantities obtained in these regions. 
Even without assuming a specific flavor structure, experimental constraints can restrict the parameter space in a way that non-trivial correlations emerge among neutrino masses, mixing parameters, and CP phases. 
Identifying these correlations may provide clues to the underlying structure of lepton-flavor hierarchies and mixing.

\medskip

This study involves two complementary steps. 
The first is a global exploration of the high-dimensional parameter space of the Yukawa matrices and Majorana masses. 
A direct random scan is generally inefficient because the allowed parameter regions can occupy only a small fraction of the entire parameter space and may have narrow, nonlinear, or disconnected structures. 
Therefore, in light of the application of various machine learning techniques in flavor physics \cite{Harvey:2021oue, Nishimura:2020nre, Matchev:2024ash, Kawai:2024pws, Nishimura:2024apb, Nishimura:2025rsk, Nishimura:2025knz, Abu-Ajamieh:2025mjk, Kitagawa:2026eck}, we use flow matching to model the distribution of parameter points that satisfy the experimental constraints. 
By learning a continuous transformation from a simple reference distribution to the distribution of viable seesaw parameters, flow matching enables us to sample broadly from the allowed regions and investigate their global structure. 
This approach yields a large ensemble of Yukawa matrices and Majorana masses that reproduce the measured neutrino mass-squared differences and mixing angles.

\medskip

The second step is to identify correlations among the physical quantities derived from the generated parameter distribution. 
For this purpose, we employ an autoencoder, which compresses the input data into a lower-dimensional latent representation and reconstructs the original variables from it. 
If the physical quantities were mutually independent, a substantial reduction in dimensionality would generally lead to a significant loss of information. 
Conversely, accurate reconstruction from a low-dimensional latent space indicates that the observables contain redundant information and lie close to a lower-dimensional structure in the full observable space. 
Since the encoder and decoder can represent nonlinear mappings, an autoencoder is well suited to identifying relations that may not be visible in two-dimensional distributions or through linear correlation coefficients. 
By analyzing the reconstruction of individual observables from the compressed representation, we search for non-trivial nonlinear relations among neutrino masses, mixing parameters, and CP phases.

\medskip

This paper is organized as follows. 
In Sec.~\ref{sec:background}, we review the Type-I seesaw mechanism and the neutrino observables used in our analysis. 
In Sec.~\ref{sec:parameter_optimization_with_flow_matching}, we introduce the flow-matching posterior estimation framework and present the parameters generated to reproduce the experimental results. 
In Sec.~\ref{sec:correlation_discovery_via_autoencoder}, we describe the autoencoder analysis and extract correlations from the learned latent space. 
Finally, Sec.~\ref{sec:con} summarizes our conclusions and discusses future directions. 
Additionally, Appendix \ref{app:flow_matching} provides a brief formulation of flow matching.

\section{Background}
\label{sec:background}

We consider Majorana neutrino masses generated through the Type-I seesaw mechanism. 
In the flavor basis where the charged-lepton Yukawa matrix is diagonal, the terms relevant to neutrino masses are written as follows:
\begin{equation}
  \mathcal{L}
  = Y^{\nu}_{i\alpha}\,\bar{N}_{i}\,\ell_{\alpha}H
  - \frac{1}{2}M_{ij}\,\bar{N}_{i}\,\bar{N}_{j}
  + \mathrm{h.c.},
  \label{eq:lagrangian}
\end{equation}
where $\ell_{\alpha}$ is a left-handed lepton doublet, $H$ is the Higgs doublet, and $N_i$ denotes a right-handed neutrino. 
The index $\alpha=e,\mu,\tau$ labels the three lepton flavors. 
The matrices $Y^{\nu}$ and $M$ are the neutrino Yukawa matrix and the Majorana mass matrix, respectively.

\medskip

At energies well below the right-handed neutrino mass scale, the heavy fields can be removed from the low-energy description. 
This procedure induces the light-neutrino mass matrix
\begin{equation}
  (m_{\nu})_{\alpha\beta}
  = \langle H\rangle^{2}
    (Y^{\nu})_{\alpha i}
    (M^{-1})_{ij}
    (Y^{\nu})_{j\beta},
  \label{eq:seesaw}
\end{equation}
with the Higgs vacuum expectation value $\langle H\rangle$. 
For the numerical analysis, we work in the charged-lepton mass basis and select a basis in which the Majorana mass matrix is diagonal.
The matrix $m_{\nu}$ is complex and symmetric, and can be diagonalized by a unitary matrix $U_{\mathrm{PMNS}}$, 
\begin{equation}
  U_{\mathrm{PMNS}}^{T} m_{\nu} U_{\mathrm{PMNS}}
  = \mathrm{diag}(m_1,m_2,m_3),
  \label{eq:takagi}
\end{equation}
where the eigenvalues $m_i$ are chosen to be non-negative. 
Normal ordering (NO) of neutrino masses corresponds to $m_1 < m_2 < m_3$, whereas inverted ordering (IO) is characterized by $m_3 < m_1 < m_2$. 
This work focuses on the NO case, but the same analytical method is applicable for the IO case.

\medskip

A convenient parameterization of the leptonic mixing matrix separates the three mixing angles, one Dirac phase, and two Majorana phases:
\begin{align}
\begin{split}
    U_{\mathrm{PMNS}}
  ={}&
  \begin{pmatrix}
    c_{12}c_{13} & s_{12}c_{13} & s_{13}e^{-i \delta_{\mathrm{CP}}} \\
    -s_{12}c_{23}-c_{12}s_{23}s_{13}e^{i \delta_{\mathrm{CP}}}
      & c_{12}c_{23}-s_{12}s_{23}s_{13}e^{i \delta_{\mathrm{CP}}}
      & s_{23}c_{13} \\
    s_{12}s_{23}-c_{12}c_{23}s_{13}e^{i \delta_{\mathrm{CP}}}
      & -c_{12}s_{23}-s_{12}c_{23}s_{13}e^{i \delta_{\mathrm{CP}}}
      & c_{23}c_{13}
  \end{pmatrix} \\
  &\times
  \begin{pmatrix}
    1 & 0 & 0 \\
    0 & e^{i\alpha_{21}/2} & 0 \\
    0 & 0 & e^{i\alpha_{31}/2}
  \end{pmatrix},
  \label{eq:pmns-parametrization}
\end{split}
\end{align}
where $c_{ij}=\cos\theta_{ij}$ and $s_{ij}=\sin\theta_{ij}$.

\medskip

The Dirac CP phase $\delta_{\mathrm{CP}}$ is encoded in the rephasing-invariant quantity known as the Jarlskog invariant:
\begin{equation}
  J_{\mathrm{CP}}
  = \operatorname{Im}\!\left(U_{e1}U_{\mu2}U_{e2}^{*}U_{\mu1}^{*}\right)
  = s_{23}c_{23}s_{12}c_{12}s_{13}c_{13}^{2}\sin \delta_{\mathrm{CP}}.
  \label{eq:jarlskog}
\end{equation}
with $U_{\alpha i}=\left(U_{\mathrm{PMNS}}\right)_{\alpha i}$. 
Information on the two Majorana phases $\alpha_{21}, \alpha_{31}$ may likewise be expressed through the following invariants.
\begin{align}
  I_1
  &= \operatorname{Im}\!\left(U_{e1}^{*}U_{e2}\right)
   = c_{12}s_{12}c_{13}^{2}\sin\!\left(\frac{\alpha_{21}}{2}\right),
  \label{eq:majorana-invariant-1}\\
  I_2
  &= \operatorname{Im}\!\left(U_{e1}^{*}U_{e3}\right)
   = c_{12}s_{13}c_{13}
     \sin\!\left(\frac{\alpha_{31}}{2}-\delta_{\mathrm{CP}} \right).
  \label{eq:majorana-invariant-2}
\end{align}

\medskip

Table \ref{tab:NuFIT} presents the experimental results for the observables in the lepton sector, as reported by NuFIT 6.0 \cite{Esteban:2024eli}.
According to this, the magnitudes of the PMNS matrix elements lie within the following 3$\sigma$ C.L.~intervals.
\begin{equation}
  |U_{\mathrm{PMNS}}|_{3\sigma}
  =
  \begin{pmatrix}
    0.801 \rightarrow 0.842, & 0.519 \rightarrow 0.580, & 0.142 \rightarrow 0.155 \\
    0.252 \rightarrow 0.501, & 0.496 \rightarrow 0.680, & 0.652 \rightarrow 0.756 \\
    0.276 \rightarrow 0.518, & 0.485 \rightarrow 0.673, & 0.637 \rightarrow 0.743
  \end{pmatrix}.
  \label{eq:pmns-3sigma}
\end{equation}

\medskip

Cosmological constraints on the sum of neutrino masses have been reported by DESI \cite{DESI:2024mwx}. 
The CMB dataset, which includes the primary Planck temperature and polarization measurements combined with Planck–ACT lensing data, yields the following 95\% C.L.~limit under the assumption of a spatially flat $\Lambda$CDM cosmology: 
\begin{align}
\sum m_\nu < 210\,\meV.
\end{align}
When DESI BAO data are added to the same CMB combination, the allowed neutrino mass range is significantly reduced, resulting in 
\begin{align}
\sum m_\nu < 72\,\meV, \label{eq:DESI_BAO+CMB}
\end{align}
at the same confidence level. 
This upper bound depends on the prior assumed for the total neutrino mass. 
While Eq.~\eqref{eq:DESI_BAO+CMB} is derived assuming the prior $\sum m_\nu >0\,\eV$, one can also consider the NO case ($\sum m_\nu >0.059\,\eV$) and the IO case ($\sum m_\nu >0.10\,\eV$). 
Here, these lower bounds are determined by NuFIT. 
For the NO scenario, the DESI BAO plus CMB analysis modifies the 95\% upper bound as follows: 
\begin{align}
\sum m_\nu < 113\,\meV. \label{eq:DESI_BAO+CMB_NO}
\end{align}
For the IO scenario, the corresponding constraint becomes $\sum m_\nu < 145\,\meV$.

\medskip

For neutrinoless double-beta decay, the effective Majorana neutrino mass is defined as follows: 
\begin{equation}
  \langle m_{ee}\rangle
  = \left|
      m_1c_{12}^{2}c_{13}^{2}
      +m_2s_{12}^{2}c_{13}^{2}e^{i\alpha_{21}}
      +m_3s_{13}^{2}e^{i(\alpha_{31}-2\delta_{\mathrm{CP}})}
    \right|.
  \label{eq:mbb}
\end{equation}
KamLAND-Zen \cite{KamLAND-Zen:2022tow} reports an upper bound of $\langle m_{ee}\rangle<36\,\mathrm{meV}$ at $90\%$ C.L.. 
Regarding future experiments, the sensitivity of the nEXO experiment~\cite{nEXO:2021ujk} is represented by the following exclusion band.
\begin{align}\label{eq:nEXO}
    4.7 \leq \langle m_{ee}\rangle/\meV \leq 20.3.
\end{align}

\medskip

\renewcommand{\arraystretch}{1.25}
\begin{table}[th]
\caption{The experimental results adopted in this work are taken from the NuFIT 6.0 global fit incorporating the Super-Kamiokande atmospheric-neutrino data \cite{Esteban:2024eli}. For neutrino masses, $\Delta m_{3l}^{2}\equiv \Delta m_{31}^{2}=m^2_3 -m^2_1>0$ for NO and $\Delta m_{3l}^{2}\equiv \Delta m_{32}^{2}=m^2_3 -m^2_2<0$ for IO.}
\label{tab:NuFIT}
\centering
\small
   \begin{tabular}{|c||c|c||c|c|} \hline
     \multirow{2}{*}{Observables} & \multicolumn{2}{c||}{Normal Ordering (NO)} & \multicolumn{2}{c|}{Inverted Ordering (IO)}  \\
     \cline{2-5}
       & best fit $\pm 1\sigma$ & $3\sigma$ range & best fit $\pm 1\sigma$ & $3\sigma$ range  \\
     \hline
     $\sin^{2}\theta_{12}$ & $0.308_{-0.011}^{+0.012}$ & $0.275\rightarrow 0.345$ & $0.308_{-0.011}^{+0.012}$ & $0.275\rightarrow 0.345$ \\
     $\theta_{12}/\tcdegree$ & $33.68_{-0.70}^{+0.73}$ & $31.63\rightarrow 35.95$ & $33.68_{-0.70}^{+0.73}$ & $31.63\rightarrow 35.95$ \\
     \hline
     $\sin^{2}\theta_{13}$ & $0.02215_{-0.00058}^{+0.00056}$ & $0.02030\rightarrow 0.02388$ & $0.02231_{-0.00056}^{+0.00056}$ & $0.02060\rightarrow 0.02409$ \\
     $\theta_{13}/\tcdegree$ & $8.56_{-0.11}^{+0.11}$ & $8.19\rightarrow 8.89$ & $8.59_{-0.11}^{+0.11}$ & $8.25\rightarrow 8.93$ \\
     \hline
      $\sin^{2}\theta_{23}$ & $0.470_{-0.013}^{+0.017}$ & $0.435\rightarrow 0.585$ & $0.550_{-0.015}^{+0.012}$ & $0.440\rightarrow 0.584$ \\
      $\theta_{23}/\tcdegree$ & $43.3_{-0.8}^{+1.0}$ & $41.3\rightarrow 49.9$ & $47.9_{-0.9}^{+0.7}$ & $41.5\rightarrow 49.8$ \\
     \hline
      $\delta_{\text{CP}}/\pi$ & $1.18_{-0.23}^{+0.14}$ & $0.69\rightarrow 2.02$ & $1.52_{-0.14}^{+0.12}$ & $1.12\rightarrow 1.86$ \\
      $\delta_{\text{CP}}/\tcdegree$ & $212_{-41}^{+26}$ & $124\rightarrow 364$ & $274_{-25}^{+22}$ & $201\rightarrow 335$ \\
     \hline
     $\Delta m_{21}^{2}/10^{-5}\,\eV^{2}$ & $7.49_{-0.19}^{+0.19}$ & $6.92\rightarrow 8.05$ & $7.49_{-0.19}^{+0.19}$ & $6.92\rightarrow 8.05$ \\
     \hline
     $\Delta m_{3l}^{2}/10^{-3}\,\eV^{2}$ & $+2.513_{-0.019}^{+0.021}$ & $+2.451\rightarrow +2.578$ & $-2.484_{-0.020}^{+0.020}$ & $-2.547\rightarrow -2.421$ \\ 
     \hline
 \end{tabular}
 \renewcommand{\arraystretch}{1}
\end{table}

\section{Parameter Optimization with Flow Matching}
\label{sec:parameter_optimization_with_flow_matching}

Determining the seesaw parameters from measured neutrino observables constitutes an inverse problem. 
Although the forward mapping from the model parameters $G$ to the observables $L$ can be evaluated straightforwardly using the Type-I seesaw relation, the inverse mapping is generally non-unique and cannot be expressed in closed form. 
Therefore, we formulate the parameter search as a simulation-based inference problem, in which samples drawn from a prescribed prior distribution over $G$ are propagated through the forward model and subsequently used to learn the conditional distribution $p(G|L)$. 
In particular, flow matching posterior estimation provides a continuous normalizing flow approach to conditional density estimation and is well suited to inverse problems with high-dimensional and multimodal posteriors.

\medskip

Indeed, Refs.~\cite{Nishimura:2025rsk, Nishimura:2025knz, Kitagawa:2026eck} are cited as previous studies utilizing generative AI. 
Specifically, flow matching has already been employed in Ref.~\cite{Kitagawa:2026eck} as a conditional generative sampler to identify phenomenologically viable regions of neutrino-texture parameter spaces, thereby demonstrating its practical utility in flavor-physics analyses. 
In the present study, we extend this line of research by revisiting the underlying formulation of flow matching and more concretely demonstrating how data-driven inference can be used to explore and characterize nontrivial flavor structures, while also improving the efficiency of collecting informative data.

\subsection{Method}
\label{sec:method_FM}

Let $G$ denote the parameters to be fitted and $L$ the calculated observables. 
Specifically, these are defined as follows:
\begin{align}
\begin{split}
    G &= \left\{\Re Y^{\nu}_{i\alpha}, \Im Y^{\nu}_{i\alpha},\frac{M_1}{M_3}, \frac{M_2}{M_3}, \log_{10}\left(\frac{\langle H\rangle^2}{M_3}/\meV\right), \arg M_1, \arg M_2\right\}, \\
    L &= \left\{\log_{10}\left(\Delta m_{21}^2/\meV^2\right), \log_{10}\left(\Delta m_{31}^2/\meV^2\right),|\left(U_{\mathrm{PMNS}}\right)_{\alpha i}|\right\}.
\end{split}
\end{align}
Since $Y^\nu$ is taken to be a general complex $3\times3$ matrix, its real and imaginary parts provide 18 real parameters. 
Together with the two heavy-mass ratios, the logarithmic scale parameter, and the two phases, the parameter vector has dimension $d_G=23$. 
The observable vector contains two mass-squared differences and the nine absolute values of the PMNS matrix elements, and hence has dimension $d_L=11$. 
The following ranges are adopted for each element of $G$.
\begin{align}
\begin{split}
    -1 \leq \left\{ \Re Y_{i\alpha}^{\nu},\,\Im Y_{i\alpha}^{\nu} \right\} \leq 1, &\quad
    10^{-5} \leq \left\{ \frac{|M_{1}|}{M_{3}},\,\frac{|M_{2}|}{M_{3}} \right\} \leq 1, \\
    -\pi \leq \left\{ \arg M_{1},\,\arg M_{2} \right\} \leq \pi, &\quad
    -1.22 \leq \log_{10} \left(\frac{\langle H \rangle ^2}{M_{3}}/\meV\right) \leq 2.78.
\end{split}
\label{eq:G_range}
\end{align}
The last is equivalent to $ 6.05\times10^{-2}\,\meV \leq \langle H \rangle ^2 / M_{3} \leq 6.05\times10^{2}\,\meV $, which is derived from $10^{14}\,\GeV \leq M_{3} \leq 10^{18}\,\GeV$. 
The neutrino Yukawa couplings and right-handed neutrino mass parameters used to construct the training dataset are randomly sampled from the ranges specified in Eq.~\eqref{eq:G_range}. 
This approach is conceptually similar to neutrino mass anarchy~\cite{Hall:1999sn}, where flavor structures are explored statistically without assuming a specific flavor symmetry or texture. 
However, in this work, the randomly sampled parameters serve as training data for flow-matching posterior estimation, enabling us to identify the subset of the parameter space consistent with the experimentally measured neutrino observables. 
We emphasize that the distributions obtained in this analysis should be interpreted as properties of the experimentally conditioned ensemble, given the parameter ranges and sampling measure defined in Eq.~\eqref{eq:G_range}. 
Therefore, they are not intended to represent prior-independent predictions of the most general Type-I seesaw mechanism. 
Instead, our objective is to investigate whether non-trivial structures emerge after imposing experimental constraints on a broad, texture-independent ensemble of high-energy parameters.

\medskip

Although the leptonic mixing pattern can be described by a minimal set of three mixing angles along with the CP phases, we use the nine absolute values of the PMNS matrix elements as conditioning variables. 
These nine quantities are not independent; however, in our numerical tests, conditioning the flow-matching model on $|(U_{\rm PMNS})_{\alpha i}|$ resulted in more accurate generation than conditioning it directly on the three mixing angles. 
Therefore, using all nine elements is a numerical choice made to improve the stability and accuracy of the conditional density estimation.
\footnote{In addition, the absolute values of the PMNS matrix elements are independent of the Majorana phases $\alpha_{21}, \alpha_{31}$. 
The phase structures observed later in Sec.~\ref{sec:correlation_discovery_via_autoencoder} thus arise indirectly from the distribution of free parameters constrained by experimental data. 
This context explains why we interpret the phase correlations as intrinsic properties of the generated parameter ensemble.}

\medskip

When $L_{\mathrm{exp}}$ represents the experimental value of $L$, it is challenging to directly determine the parameter $G$ that reproduces $L_{\mathrm{exp}}$. 
Therefore, we formulate the parameter optimization as the task of estimating the posterior distribution $p(G|L_{\mathrm{exp}})$, which represents the probability that $G$ yields $L_{\mathrm{exp}}$ given the observed $L_{\mathrm{exp}}$. 
By sampling parameters from $p(G|L_{\mathrm{exp}})$, it becomes possible to efficiently search for parameters that actually yield values close to $L_{\mathrm{exp}}$. 
As a technique for estimating this posterior distribution, we employ a method called Flow Matching Posterior Estimation (FMPE) \cite{Dax:2023ozk}, which is a type of generative model.
The basic formulation of flow matching
\footnote{Since its inception, flow matching is extended in various directions. 
For example, Discrete Flow Matching \cite{Gat:2024dfm}, Flow Map Matching \cite{Boffi:2025flo}, and MeanFlow \cite{Geng:2025mea} have been proposed; these methods directly learn flow maps or average velocities rather than velocity fields. 
In particular, the latter aims to replace the sequential integration of ODEs with one-step or few-step mappings, representing a significant advancement toward accelerating the generation process. 
Furthermore, LieFlow \cite{Chen:2026lie} has been developed as a method to learn and discover continuous or discrete symmetries inherent in the data. 
While this study employs the standard FMPE, these developments are considered valuable for accelerating inference and for incorporating the symmetries of physical systems into generative AI.}
is explained in the Appendix \ref{app:flow_matching}. 
Here, the term ``posterior'' refers to the conditional distribution induced by the simulation prior and the specified observation $L_{\rm exp}$ within the FMPE framework. 
It should not be confused with a full Bayesian posterior constructed directly from the experimental likelihood, which includes all experimental correlations and systematic uncertainties.

\medskip

We briefly introduce the procedures of FMPE. 
Since $p(G|L)$ is generally complex, FMPE employs the following approach. 
First, the flow $\psi_{t, L}(G)$ is defined by a certain vector field $u_{t,L}(G)$:
\begin{equation}
    \frac{d}{dt}\psi_{t, L}(G) = u_{t,L}(\psi_{t, L}(G)).
\end{equation}
The following boundary conditions are imposed. 
\begin{equation}\label{eq:p_path_marginal_conditions}
    p_{0, L}(G) = p(G),\quad p_{1, L}(G) = p(G|L).
\end{equation}
Note that the time variable $t\in[0,1]$ parameterizes the probability path and is not a physical time. 
Next, consider obtaining $p(G|L)$ by transforming the normal distribution $p(G)=\mathcal{N}(G|0, I)$ via the flow $\psi_{t, L}(G)$ defined by the vector field $u_{t,L}(G)$: 
\begin{align}
\begin{split}
    p_{t, L}(G) &= [\psi_{t, L}]_*p(G) \\
    &= p(\psi_{t,L}^{-1}(G))\det\left(\frac{\partial\psi_{t,L}^{-1}}{\partial G}(G)\right). \label{eq:def_p_by_flow}
\end{split}
\end{align}
When Eq.~\eqref{eq:def_p_by_flow} holds, the vector field $u_{t, L}$ is said to generate a probability path $p_{t, L}$. 
Such a vector field is learned using another vector field $v_{t, L}(G;\theta)$, which has learning parameters $\theta$. 
It has been shown that a loss function designed as follows can achieve sufficient learning accuracy (see Ref.~\cite{Dax:2023ozk}). 
\begin{equation}\label{eq:loss_fmpe}
    \mathcal{L}_{\mathrm{FMPE}}=\mathbb{E}_{t\sim p(t), (G_1, L_1) \sim p(L_1|G_1)p(G_1), G_t\sim p_t(G_t|G_1)} ||v_{t,L_1}(G_t;\theta)-u_t(G_t|G_1)||^2,
\end{equation}
with
\begin{align}
    p_t(G|G_1) &= \mathcal{N}(G|\mu_t(G_1),\sigma_t^2(G_1)), \label{eq:def_conditional_p_path} \\
    u_{t, L}(G|G_1) &= \frac{\sigma_t^{\prime}(G_1)}{\sigma_t(G_1)}(G-\mu_t(G_1))+\mu_t^{\prime}(G_1).
\end{align}
Here, $\mu_t$ and $\sigma_t$ are functions that satisfy the following conditions, where $\sigma_{\min}$ is a small positive value.
\begin{align}
\begin{split}
    \mu_0(G) = 0,&\quad \mu_1(G) = G, \\
    \sigma_{0}(G) = 1,&\quad \sigma_1(G)=\sigma_{\min}.
\end{split}
\end{align}
As a commonly used choice, this study adopts
\begin{equation}
    \mu_t(G) = tG,\quad\sigma_t(G)=1-(1-\sigma_{\min})t,
\end{equation}
which derives the following expression. 
In practice, $\sigma_{\min}\to 0$ is adopted.
\begin{equation}\label{eq:conditional_u_t}
    u_t(G|G_1) = \frac{G_1 - (1-\sigma_{\min})G}{1-(1-\sigma_{\min})t}.
\end{equation}

\medskip

$(G_1, L_1) \sim p(L_1|G_1)p(G_1)$ can be collected during training by sampling a large number of $G_1$ instances from a uniform distribution and then computing $L_1$ for each $G_1$ using the Type-I seesaw mechanism. 
The training procedure is summarized below.
\begin{enumerate}
    \item \textbf{Simulation}

    We sample a large number of values of $G_1$ from a uniform distribution. 
    Then, for each $G_1$, the observable $L_1$ is calculated using the Type-I seesaw mechanism. 
    This process produces the training dataset $\mathcal{D}$:
    \begin{equation}
        \mathcal{D}=\left\{\left(G_1^{(n)},L_1^{(n)}\right)\right\}_{n=1}^{N}.
    \end{equation}

    \item \textbf{Training}

    Using the training data $\mathcal{D}$, we train the conditional vector field $v_{t,L}(G;\theta)$. 
    At each step, we sample $t \sim p(t)$ and draw $G_t$ from the conditional probability path defined in Eq.~\eqref{eq:def_conditional_p_path}. 
    Then, using the vector field in Eq.~\eqref{eq:conditional_u_t} as the teaching signal, we optimize $\theta$ to minimize the loss function given in Eq.~\eqref{eq:loss_fmpe}.

    \item \textbf{Generation}

    In this phase, the condition is fixed as the experimental value $L_{\mathrm{exp}}$. 
    First, an initial point $G_0 \sim \mathcal{N}(0, I)$ is sampled from the standard normal distribution. 
    Next, we numerically solve the following ordinary differential equation, governed by the trained vector field $v_{t,L_{\mathrm{exp}}}(G;\theta)$, from $t=0$ to $t=1$.
    \begin{equation}
        \frac{dG_t}{dt} = v_{t,L_{\mathrm{exp}}}(G_t;\theta).
    \end{equation}
    Finally, we use the resulting $G_1$ as a sample from the posterior distribution $p(G|L_{\mathrm{exp}})$.

    \item \textbf{Evaluation}

    The set of observables calculated from the parameter $G_1$ is defined as follows:
    \begin{equation}
        T = \left\{\log_{10}\left(\Delta m_{21}^2/\meV^2\right), \log_{10}\left(\Delta m_{31}^2/\meV^2\right), \sin^2\theta_{12}, \sin^2\theta_{13}, \sin^2\theta_{23}\right\}.
    \end{equation}
    Let $T_1$ denote the value of $T$ calculated from $G_1$, $T_{\mathrm{exp}}$ represent the best-fit value of $T$, and $3\sigma_{i}$ denote the $3\sigma$ range for each component of $T$. 
    The accuracy of $G_1$ is evaluated using the following $\chi^2$ value: 
    \begin{equation}\label{eq:def_chi2}
        \chi^2 = 9 \times\sum_{i} \frac{\left(T_1 - T_{\mathrm{exp}}\right)^2}{\left(3\sigma_i\right)^2}.
    \end{equation}
    A smaller $\chi^2$ value indicates higher accuracy of $T_1$.
\end{enumerate}

\medskip

Because the forward map $G\mapsto L$ is generally non-injective, conditioning on the experimental observables $L_{\exp}$ does not uniquely determine the underlying parameter point. 
Instead, the conditional distribution $p(G|L_{\exp})$ may be multimodal and exhibit extended degeneracy directions. 
Accordingly, the objective of the generative model is not to identify a single best-fit solution, but to generate a representative ensemble of parameter points that captures the full structure of these degeneracies.

\subsection{Network Architecture}

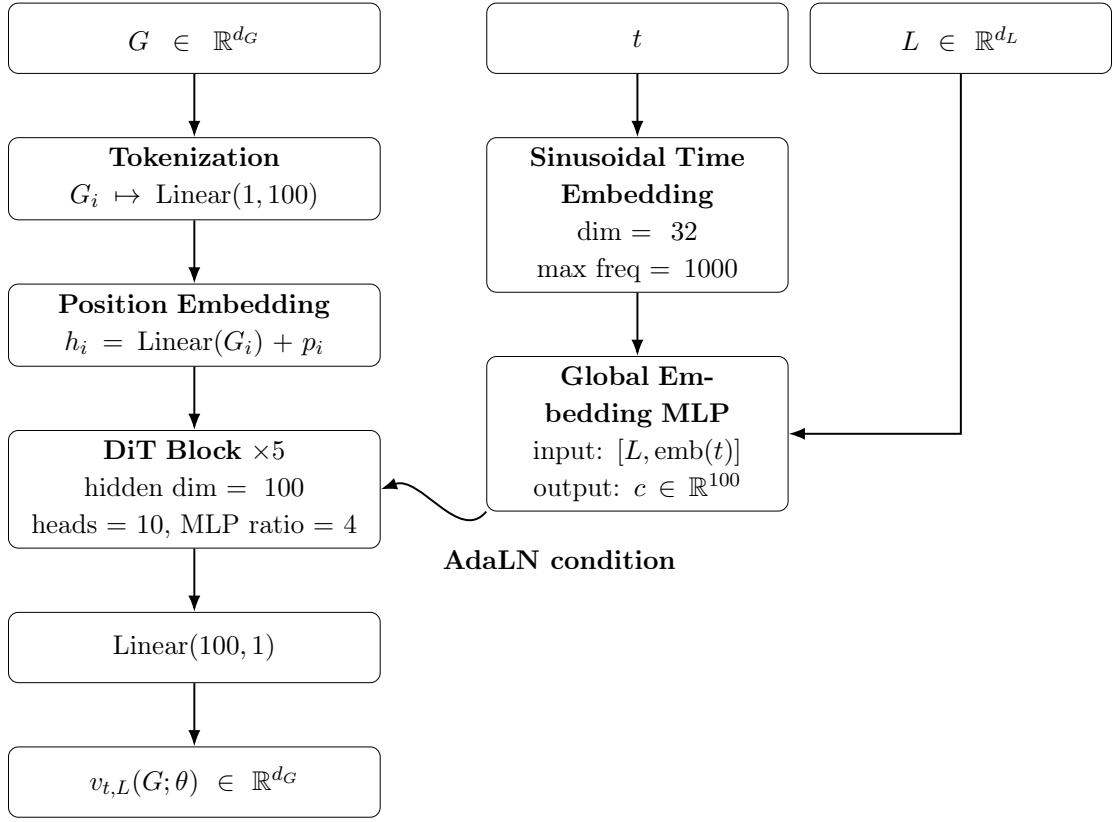
\begin{figure}[th]
\centering
\resizebox{0.98\linewidth}{!}{
\begin{tikzpicture}[
    block/.style={
        rectangle,
        draw,
        rounded corners,
        align=center,
        text width=5.0cm,
        minimum height=1.0cm,
        inner sep=4pt
    },
    smallblock/.style={
        rectangle,
        draw,
        rounded corners,
        align=center,
        text width=4.0cm,
        minimum height=1.0cm,
        inner sep=4pt
    },
    arrow/.style={-{Latex}, thick},
    node distance=0.9cm
]

\node[block] (G) {$G \in \mathbb{R}^{d_G}$};

\node[block, below=of G] (token) {
\textbf{Tokenization}\\
$G_i \mapsto \operatorname{Linear}(1,100)$
};

\node[block, below=of token] (pos) {
\textbf{Position Embedding}\\
$h_i=\operatorname{Linear}(G_i)+p_i$
};

\node[block, below=of pos] (dit) {
\textbf{DiT Block} $\times 5$\\
hidden dim $=100$\\
heads $=10$, MLP ratio $=4$
};

\node[block, below=of dit] (outproj) {
$\operatorname{Linear}(100,1)$
};

\node[block, below=of outproj] (vf) {
$v_{t,L}(G;\theta) \in \mathbb{R}^{d_G}$
};

\node[smallblock, right=1.5cm of G] (t) {$t$};

\node[smallblock, below=of t] (temb) {
\textbf{Sinusoidal Time Embedding}\\
dim $=32$\\
max freq $=1000$
};

\node[smallblock, right=0.3cm of t] (L) {
$L \in \mathbb{R}^{d_L}$
};

\node[smallblock, below=of temb] (global) {
\textbf{Global Embedding MLP}\\
input: $[L,\operatorname{emb}(t)]$\\
output: $c \in \mathbb{R}^{100}$
};

\draw[arrow] (G) -- (token);
\draw[arrow] (token) -- (pos);
\draw[arrow] (pos) -- (dit);
\draw[arrow] (dit) -- (outproj);
\draw[arrow] (outproj) -- (vf);

\draw[arrow] (t) -- (temb);
\draw[arrow] (temb) -- (global);
\draw[arrow] (L.south) |- (global.east);

\draw[arrow] (global.south west) to[out=230,in=20]
node[pos=0.55, below right, yshift=-15pt] {\textbf{AdaLN condition}} (dit.east);

\end{tikzpicture}
}
\caption{Transformer architecture for vector field estimation.}
\label{fig:transformer_network}
\end{figure}

In this study, we use version 0.25.0 of the Python library \textit{sbi}, as provided in Ref.~\cite{BoeltsDeistler_sbi_2025}. 
Although this library offers several options for the neural network architecture of the vector field $v_{t, L}(G;\theta)$, we employ the Transformer architecture \cite{Vaswani:2017lxt, William:2023sca}. 
This choice is expected not only to enable the individual processing of each component of $G$, but also to capture the dependencies between the components of the Yukawa matrix and the Majorana mass, while modulating the output according to the observation conditions $L$ and the time $t$ along the probability path. 
The specific architecture is illustrated in Fig.~\ref{fig:transformer_network}.

\medskip

The roles of each component are as follows. 
In the \textit{tokenization}, each component $G_i$ of the input vector $G$ is transformed into a 100-dimensional token representation using $\mathrm{Linear}(1,100)$. 
This process embeds each parameter, originally a single real number, into a common feature space that can be effectively processed by self-attention mechanisms and multilayer perceptrons (MLPs).
In the \textit{position embedding}, a position embedding $p_i$ is added to each token to provide positional information for each component of the input vector. 
Since the self-attention mechanism cannot inherently distinguish the order of tokens without positional information, it is necessary to explicitly specify which token corresponds to which physical quantity. 
Through this process, for example, the model can differentiate components corresponding to Yukawa matrix elements from those corresponding to Majorana masses, thereby learning the dependencies between them.
In the \textit{sinusoidal time embedding}, the time $t$ is transformed into a 32-dimensional periodic embedding vector $\mathrm{emb}(t)$. 
Since the appropriate vector field varies depending on the position along the stochastic path, providing the network with information about $t$ enables it to represent transformations suitable for each stage of the transport process, from start to finish.
In the \textit{global embedding MLP}, a 100-dimensional conditional embedding $c$ is generated using the conditional vector $L$ and the time embedding $\mathrm{emb}(t)$.
In the \textit{diffusion transformer (DiT) block}, the conditional embedding $c$ is used to generate the shift, scale, and gate parameters for adaptive layer normalization. 
These parameters conditionally modulate the self-attention mechanism and the MLP, enabling the model to perform transformations that depend on the condition $L$ and the time step $t$, while learning the dependencies among the components of $G$.
Finally, the \textit{output projection} maps each token back to one dimension using $\mathrm{Linear}(100,1)$, producing a vector field $v_{t,L}(G;\theta)$ with the same dimension as the input $G$. 
Thus, the purpose of adopting the Transformer architecture is not merely to process high-dimensional inputs, but to capture the nonlinear dependencies and degenerate structures among the parameters that arise in the inverse problem of the Type-I seesaw mechanism, representing them as a vector field controlled by the observation condition $L$ and the time $t$.

\medskip

In principle, the choice of a Transformer-based architecture is not unique; other network architectures can also be used to represent the conditional vector field. 
Nevertheless, in our numerical comparisons, the current Transformer architecture achieved higher generation accuracy than a simple fully connected network under otherwise comparable training settings. 
This improvement is attributed to the self-attention mechanism, which enables the network to model dependencies among the different components of $G$, while incorporating the observation vector $L$ and the flow time $t$ through the conditional embedding. 
Therefore, we conclude that the Transformer architecture is a practical approach for improving the accuracy and stability of posterior estimation.

\medskip

\subsection{Numerical Implementation and Results}
\label{sec:result_FM}

This section explains the actual procedures for collecting an appropriate dataset, following the four stages described in Sec.~\ref{sec:method_FM}. 
All calculations are conducted with an NVIDIA GeForce RTX 4070 Laptop GPU with 8 GB of VRAM.

\subsubsection*{Simulation}
\label{sec:simulation}

Each component of $G$ is sampled as follows. 
First, $10^7$ samples $\tilde{G_i}\sim\mathcal{U}(0, 1)$ are drawn from the uniform distribution $\mathcal{U}(0, 1)$. 
These samples are then linearly transformed using the following equation so that the distribution corresponds to the range of the $i$-th component of $G$. 
\begin{equation}
    G_i = G_i^{\rm{low}} + \left(G_i^{\rm{up}} - G_i^{\rm{low}}\right) \times \tilde{G_i}.
\end{equation}
Here, $G_i^{\rm{low}}$ and $G_i^{\rm{up}}$ represent the lower and upper bounds of the range of the $i$-th component of $G$, respectively. 
In practice, they are determined by Eq.~\eqref{eq:G_range}. 
For each sample of $G$ obtained in this way, we compute $L$. 
We used a GPU for this computation, which completed in 22 seconds with a batch size of 10,000.

\newpage

The $L$ values obtained in the above process are standardized independently for each component using the mean and standard deviation of that component. In practice, this processing is performed internally by the \textit{sbi} library. 
\begin{equation}\label{eq:norm_label}
    \tilde{L_i} = \frac{L_i - \mu_i}{\sigma_i}.
\end{equation}
The normalization constants $\mu_i$ and $\sigma_i$ are computed from the training partition and then kept fixed when transforming the validation data and the experimental condition $L_{\exp}$. 
The same constants are also used when conditioning the network during generation. 
This convention prevents the introduction of information from the validation sample into the training preprocessing.

\subsubsection*{Training}

We use the pairs of $\tilde{G}$ and $\tilde{L}$ obtained in the previous stage as the training dataset. 
The dataset is divided so that $90\%$ is used for training and $10\%$ for validation. 
We apply the Adam optimizer and OneCycleLR as the learning rate scheduler, with a maximum learning rate of $5\times10^{-4}$. 
The training batch size is set to 256, and the total number of training steps is 50,000. 
Validation is performed every 1,000 steps, using 10 batches of validation data. 
Training is completed in approximately 40 minutes.

\subsubsection*{Generation}

$G$ is generated using the normalized experimental values of $L$ as input. 
In practice, this normalization is performed internally by the \textit{sbi} library using the statistics obtained from Eq.~\eqref{eq:norm_label}. 
In total, $3\times10^6$ instances are generated with a batch size of 1,024, which took approximately 11 hours.

\subsubsection*{Evaluation}
\label{sec:fmpe_evaluation}

\begin{figure}[th]
    \centering
    \includegraphics[width=0.8\linewidth]{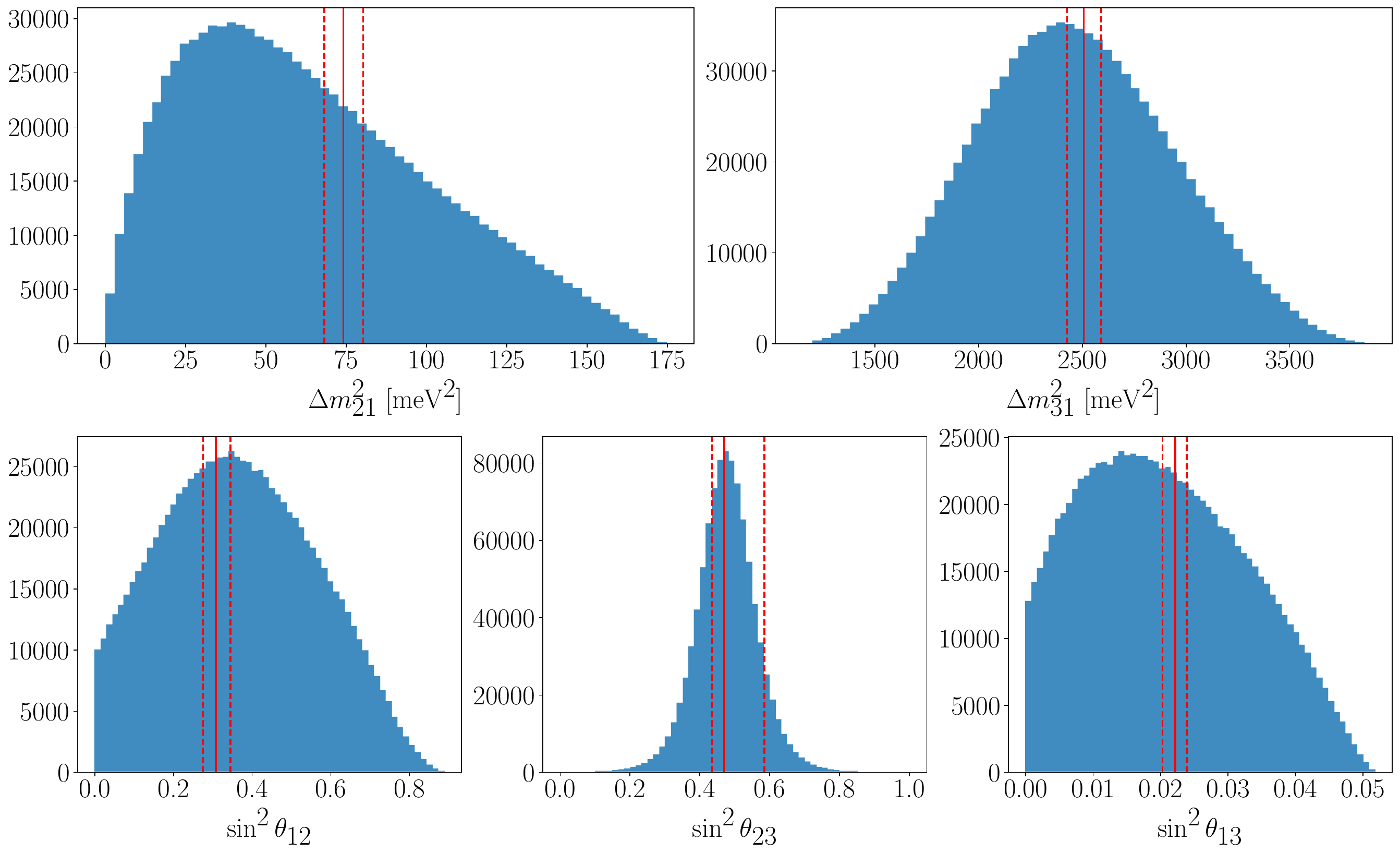}
    \caption{Results generated using flow matching. The solid red line represents the best-fit value, and the dashed line represents the 3$\sigma$ range.}
    \label{fig:generated_distributions}
\end{figure}

We compute $T$ from the generated $G$ and evaluate it using the $\chi^2$ defined in Eq.~\eqref{eq:def_chi2}. A total of 956,607 generated samples satisfy $\chi^2 < 2{,}500$, among which 640 samples satisfy the more stringent criterion of $\chi^2 < 45$. Furthermore, the distributions of the individual components of $T$ are shown in Fig.~\ref{fig:generated_distributions}. The 956,607 samples satisfying $\chi^2 < 2{,}500$ are used to train the autoencoder described in the next section.

\medskip

Interestingly, although the $10^7$ training data points used for parameter optimization via flow matching include no points with $\chi^2 < 45$, the model can discover a large number of data points with $\chi^2 < 45$ during generation.
This result demonstrates that Flow Matching does more than simply reproduce samples observed during training. 
By capturing the high-dimensional and nonlinear geometry that relates model parameters to physical observables, the model can uncover high-accuracy solutions not represented by any sample in the finite training set. 
Although further validation is required before this behavior can be classified as strict out-of-distribution extrapolation, our findings provide an important and compelling demonstration that generative AI can uncover high-precision regions in high-dimensional inverse problems and substantially expand the scope of scientific exploration.

\medskip

Regarding Fig.~\ref{fig:generated_distributions}, the absence of isolated spikes or strongly fragmented structures in the marginal distributions suggests that the generated samples cover broad, continuous regions rather than collapsing onto a small number of parameter points. 
These figures also show that the accuracy of the generated samples is not uniform across the five observables. 
For example, the distribution of $\Delta m_{31}^{2}$ is approximately centered around the experimentally preferred region, and $\sin^2 \theta_{23}$ is strongly concentrated near its best-fit value. 
In contrast, the distribution of $\Delta m_{21}^{2}$ is broader and asymmetric, with its best-fit value located away from the maximum of the generated marginal distribution. 
These differences indicate that the conditional flow reproduces the experimental conditions with varying accuracy along different observable directions. 
However, the broadness of an individual marginal distribution does not necessarily imply that the generated sample is globally inaccurate, because the criterion in Eq.~\eqref{eq:def_chi2} constrains the five observables simultaneously. 
A sample located in the tail of one marginal distribution can still yield a moderate value of $\chi^2$ if the remaining observables are sufficiently close to their preferred values.

\medskip

\section{Correlation Discovery via Autoencoder}
\label{sec:correlation_discovery_via_autoencoder}

Using the parameters with $\chi^2 < 2{,}500$ obtained in Sec.~\ref{sec:fmpe_evaluation}, we calculate the following physical observables via the Type-I seesaw mechanism.
\begin{equation}\label{eq:def_A}
    A = \left\{m_1, m_2, m_3, \sin^2\theta_{12}, \sin^2\theta_{13}, \sin^2\theta_{23}, \delta_{\rm{CP}}, \alpha_{21}, \alpha_{31}\right\}.
\end{equation}
Since these data are derived not from random parameters but from parameters that reproduce the experimental values for the mass-squared differences and mixing angles, they are expected to contain non-trivial correlations. 
However, extracting meaningful features from this nine-dimensional data by examining various combinations of physical quantities and analyzing their correlations is difficult. 
Therefore, this paper proposes a method that utilizes autoencoders.

\subsection{Method}

An autoencoder consists of an encoder $\mathcal{E}_{\theta_1}: \mathbb{R}^d\rightarrow\mathbb{R}^l$ and a decoder $\mathcal{D}_{\theta_2}:\mathbb{R}^l\rightarrow\mathbb{R}^d$. 
$\theta_1$ and $\theta_2$ are learning parameters that are trained to minimize the following loss function: 
\begin{equation}
    \mathcal{L}(\theta_1,\theta_2) = \frac{1}{Bd} \sum_{i=1}^{B}\sum_{j=1}^{d} \left( x_{ij} - \hat{x}_{ij} \right)^2, \quad \hat{x}_i = \mathcal{D}_{\theta_2}\left(\mathcal{E}_{\theta_1}(x_i)\right).
\end{equation}
Here, $B$ denotes the batch size during the training phase, and $d$ represents the dimension of the input data. 
Furthermore, the $d$-dimensional input vector at the $i$-th position in each batch is denoted by $\bm{x}_i$. 
Typically, $l$ is chosen such that $l < d$ to prevent the autoencoder from overfitting or simply learning the identity mapping. 

\medskip

The $\mathbb{R}^l$ space, which serves as the output of the encoder and the input to the decoder, is called the latent space. 
Since the decoder can only use information from the latent space, the autoencoder learns to embed the key features of the input data within this space.

\medskip

In the following, we explain a method for discovering non-trivial correlations using the latent space of an autoencoder. 
We define the encoder’s output $\bm{z}_i = \mathcal{E}_{\theta_1}(\bm{x}_i)$ for an input $\bm{x}_i$ as the latent coordinate of $\bm{x}$. 
Although the representation of $\bm{x}_i$ in terms of its latent coordinates is obtained via the decoder, it is difficult to determine this representation precisely due to reconstruction errors inherent in the decoder. 
Therefore, we consider a dataset $\{(\bm{z}_i, \bm{x}_i)\}_{i = 1, 2, \dots, N}$ to obtain the representation of $\bm{x}_i$. 
Here, $\bm{z}$ and $\bm{x}$ are treated as the explanatory and response variables, respectively. 
Applying polynomial regression or similar methods to this dataset yields the following expression: 
\begin{align}
    \bm{x} = f(\bm{z}),
\end{align}
with a function $f: \mathbb{R}^l \rightarrow \mathbb{R}^d$. 
This represents the relationship among physical quantities mediated by the $l$-dimensional components. 
Since not all information about the $d$-dimensional observables is embedded within the $l$-dimensional latent space, the accuracy of this regression may be poor for certain observables. 
Conversely, the regression accuracy is expected to improve for observables for which more information is retained in the latent space. 
Therefore, if we extract only the $d^{\prime}$ components for which the regression accuracy is high, those observables lie on an $l$-dimensional hypersurface within the $d^{\prime}$-dimensional space.

\medskip

If the physical observables were mutually independent, their joint distribution would generically occupy a finite-volume region within the high-dimensional observable space, rather than being concentrated near a low-dimensional manifold. 
Therefore, the nontrivial aspect of the autoencoder analysis is not merely that the data are mapped onto a two-dimensional latent space, since the bottleneck dimension is fixed by construction. 
Rather, it is that several physical observables can be reconstructed with high accuracy from only two latent variables and that the corresponding samples organize themselves into branch-like structures. 
Such behavior indicates that the variations of these observables are not independent, but are governed by a smaller number of collective degrees of freedom.

\subsection{Network Architecture}

In this work, the latent space dimension is fixed at two. 
As summarized in Table \ref{tab:autoencoder_architecture}, the encoder maps each input vector to a two-dimensional latent representation through fully connected layers, layer normalization, and four residual blocks. 
The residual connections stabilize the training of the nonlinear transformation and help preserve information during compression. 

\medskip

The decoder approximately mirrors the encoder and reconstructs the original observables from the latent variables. 
This two-dimensional bottleneck offers a compact representation of the correlations among the input observables and enables direct visualization of the event distribution within the latent space.

\medskip

The latent dimension $l=2$ is chosen primarily to allow direct visualization and explicit analysis of the geometry in the $(z_1,z_2)$ plane. 
This choice does not imply that the intrinsic dimension of the selected physical distribution is exactly two. 
Therefore, the relations derived below should be understood as those obtained under the assumption of a two-dimensional bottleneck. 
We confirm that a similar cluster structure emerges even when the latent space is set to three dimensions.

\begin{table}[H]
    \caption{Autoencoder architecture used to compress the observables into a two-dimensional latent space and reconstruct the original input vector.}
    \label{tab:autoencoder_architecture}
    \centering
    \begin{tabular}{|c|c|c|c|}
        \hline
        Stage & Layer & Operation & Output dimension \\
        \hline\hline
        Input & -- & Input vector & $9$ \\
        \hline
         & Encoder input & $\operatorname{Linear}(9,64) + \tanh$ & $64$ \\
         & Down projection & $\operatorname{Linear}(64,32) + \tanh$ & $32$ \\
        Encoder & Normalization & $\operatorname{LayerNorm}(32)$ & $32$ \\
         & Residual blocks & $\operatorname{ResBlock}(32) \times 4$ & $32$ \\
         & Latent projection & $\operatorname{Linear}(32,2)$ & $2$ \\
        \hline
        Latent & -- & latent space & $2$ \\
        \hline
         & Decoder input & $\operatorname{Linear}(2,32) + \tanh$ & $32$ \\
         & Normalization & $\operatorname{LayerNorm}(32)$ & $32$ \\
        Decoder & Residual blocks & $\operatorname{ResBlock}(32) \times 4$ & $32$ \\
         & Up projection & $\operatorname{Linear}(32,64) + \tanh$ & $64$ \\
         & Output projection & $\operatorname{Linear}(64,9) + \tanh$ & $9$ \\
        \hline
        Output & -- & Reconstructed vector & $9$ \\
        \hline
    \end{tabular}
\end{table}

\subsection{Numerical Implementation and Results}

This section explains the actual analysis of the latent space.
All calculations are conducted using an NVIDIA GeForce RTX 4070 Laptop GPU with 8 GB of VRAM, the same equipment used in Sec.~\ref{sec:result_FM}.

\subsubsection{Training}

The physical quantities introduced in Eq.~\eqref{eq:def_A} are normalized as follows. 
To reduce the skewness of the mass distribution while mapping the mass values onto the interval $[-1,1]$, each mass $m_i$ is transformed according to
\begin{align}\label{eq:norm_mass}
    m_i^{\prime} = -1 + 2\,\frac{\log_{10}(1 + m_i/\rm{meV}) -\ell_i^{\min}}{\ell_i^{\max} - \ell_i^{\min}},\quad (i=1,2,3).
\end{align}
Here, $\ell_i^{\min}$ and $\ell_i^{\max}$ are determined from the minimum and maximum mass values ($m_{\min}$ and $m_{\max}$) as follows:
\begin{align}
    \ell_i^{\min} = \log_{10}(1 + m_{\min}/\meV),
    \quad
    \ell_i^{\max} = \log_{10}(1 + m_{\max}/\meV).
\end{align}
The mixing angles $\sin^2\theta_{12}$, $\sin^2\theta_{13}$, and $\sin^2\theta_{23}$ are used without any transformation. 
The CP phases are normalized by dividing them by $\pi$, such that
\begin{align}\label{eq:norm_phase}
    \delta_{\rm{CP}}^\prime = \delta_{\rm{CP}} / \pi,\quad
    \alpha_{21}^\prime = \alpha_{21} / \pi,\quad
    \alpha_{31}^\prime = \alpha_{31} / \pi.
\end{align}

\medskip

From the data satisfying the condition $\chi^2 < 2{,}500$, we prepare 956,607 instances of $A$ defined by Eq.~\eqref{eq:def_A}. 
After preprocessing these instances according to Eq.~\eqref{eq:norm_mass} and Eq.~\eqref{eq:norm_phase}, we split them into 90\% training data and 10\% validation data. 
The batch size is set to 64, and the number of training steps is set to 50,000. 
A validation step is performed every 1,000 training steps, using 10 batches of validation data. 
The Adam optimizer is employed, with OneCycleLR as the learning rate scheduler, set to a maximum value of $10^{-3}$. 
The training phase is completed in approximately 7 minutes.

\subsubsection{Latent Space Analysis}

\begin{figure}[th]
    \begin{minipage}{0.49\linewidth}
        \begin{center}
        \includegraphics[height=63mm]{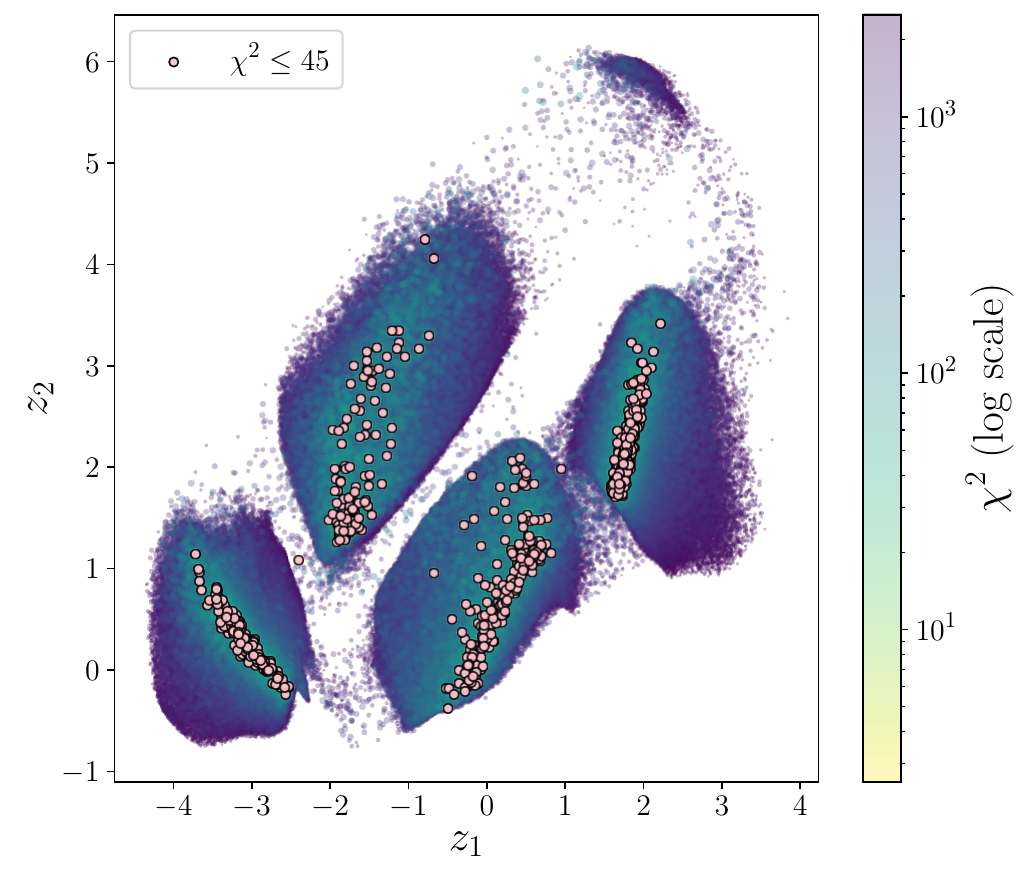}
        \end{center}
    \end{minipage}
    \begin{minipage}{0.49\linewidth}
        \begin{center}
        \includegraphics[height=64mm]{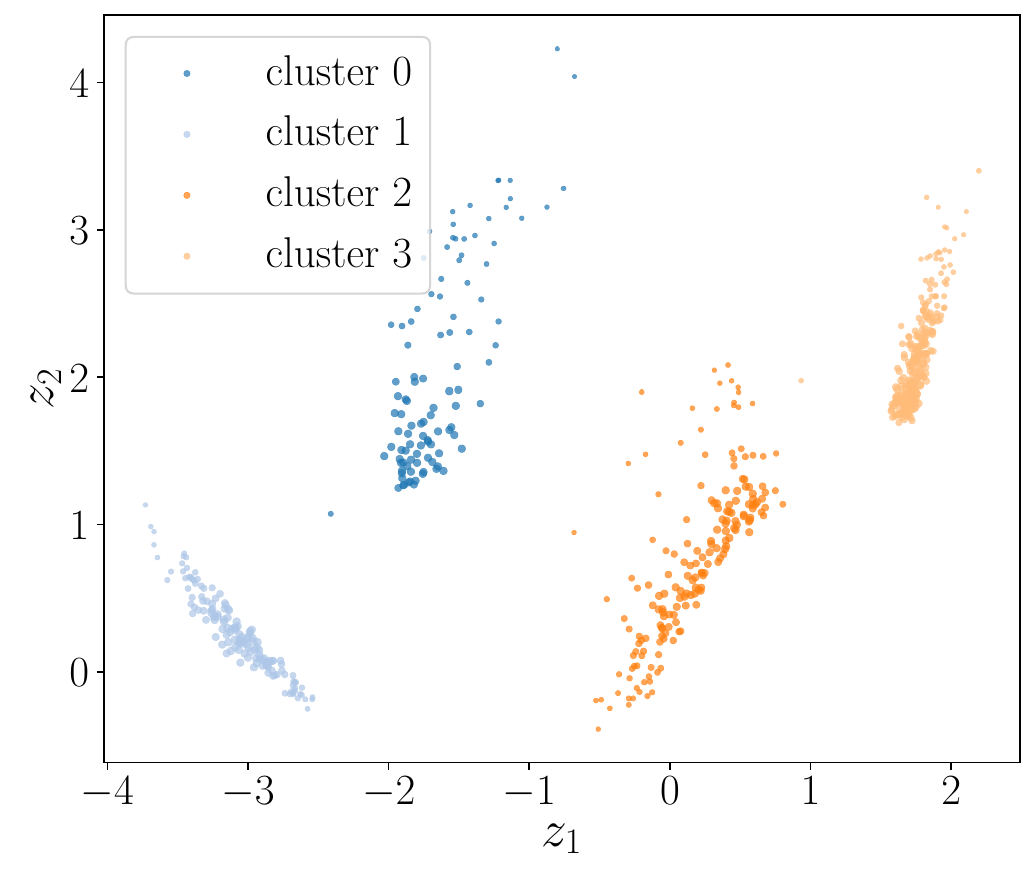}
        \end{center}
    \end{minipage}
    \caption{The left panel shows the visualization of the latent space, and the right panel shows the clustering result using HDBSCAN. Clusters 0, 1, 2, and 3 contain 105, 141, 177, and 217 points, respectively.}
    \label{fig:HDBSCAN}
\end{figure}

The left panel of Fig.~\ref{fig:HDBSCAN} shows the distribution of latent data in the bottleneck of the autoencoder.
This indicates that the encoded samples are not distributed as a single connected cloud. 
Instead, they form high-density regions in the $(z_1,z_2)$ plane.
We partition this latent space into clusters using an algorithm called HDBSCAN \cite{Campello:2013hdb}, as shown in the right panel of Fig.~\ref{fig:HDBSCAN}. 
The HDBSCAN assignment delineates the latent-space regions into clusters with distinctly different locations and orientations. 
Although the latent dimension is fixed at two by design, the emergence of several branches is not imposed by the autoencoder architecture. 
If the input observables contained no correlated structure beyond the imposed constraints, one would not generally expect the high-accuracy samples to organize into such distinct branch-like regions exhibiting cluster-dependent physical properties.
Cluster 0 occupies the upper-left branch, cluster 1 the lower-left branch, cluster 2 the central branch, and cluster 3 the upper-right branch. 
All generated samples satisfying $\chi^2<45$ are assigned to one of these four groups.

\medskip

\begin{figure}[th]
    \centering
    \includegraphics[width=\linewidth]{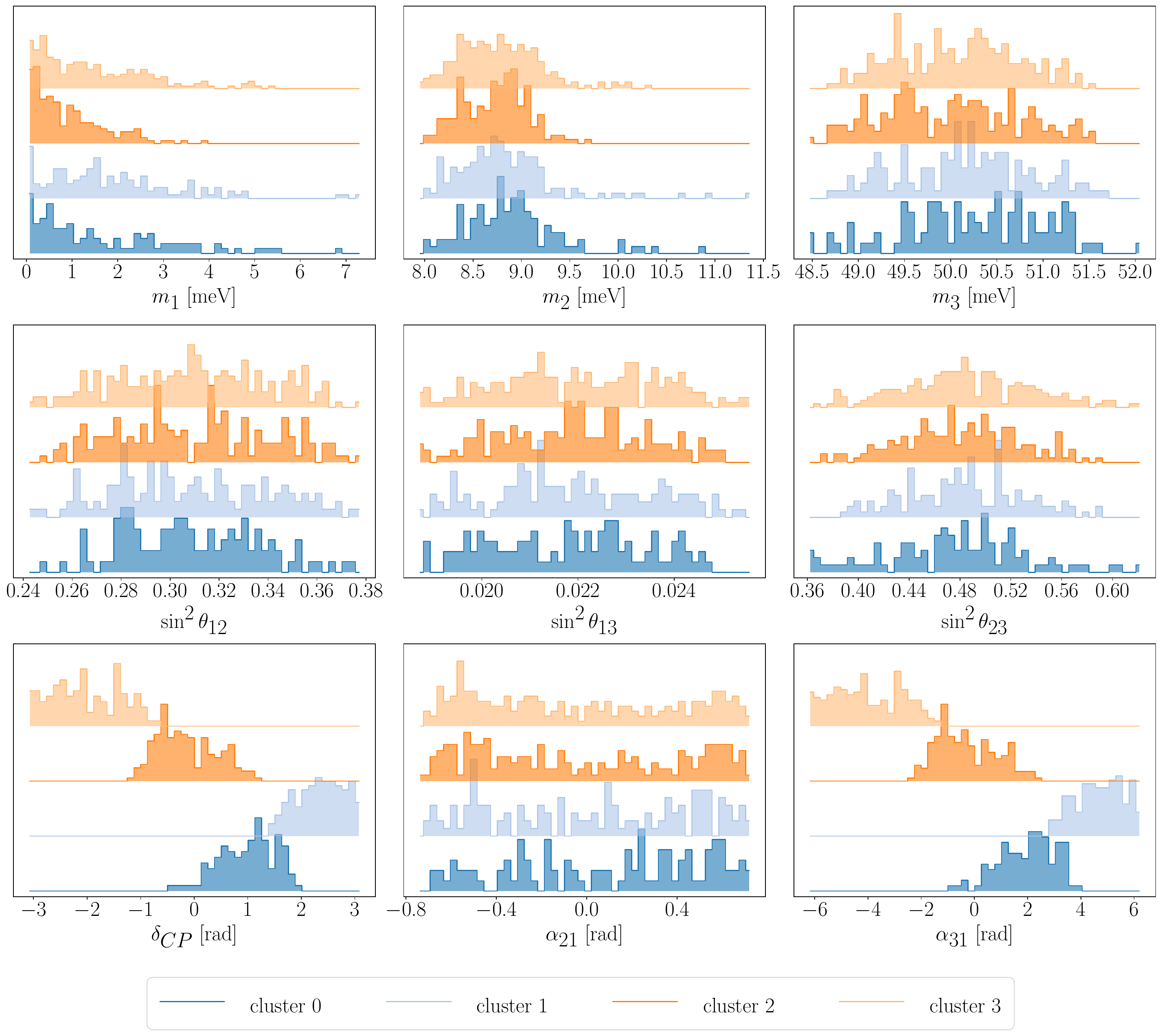}
    \caption{Distribution of observables in each cluster.}
    \label{fig:hdbscan_physical_hist}
\end{figure}

We examine the characteristics of each cluster containing points for which $\chi^2 < 45$. 
First, plotting the distributions of the observables within each cluster yields the results shown in Fig.~\ref{fig:hdbscan_physical_hist}. 
These results reveal that the most distinct cluster dependence appears in the CP phases. 
In particular, $\delta_{\rm CP}$ and $\alpha_{31}$ occupy branch-dependent intervals.

\medskip

Table \ref{tab:cluster_ranges} quantifies these trends. 
Across all clusters, $m_1$ ranges from values close to zero up to $7.36\,\meV$, whereas $m_2$ and $m_3$ remain within the narrower intervals 7.92 -- $11.4\,\meV$ and 48.4 -- $52.1\,\meV$, respectively. 
The clusters are more clearly separated by $\delta_{\rm CP}\simeq\alpha_{31}/2$: cluster 1 is restricted to positive phases between $+1.43$ -- $+3.14\,\mathrm{rad}$, cluster 3 to negative phases between $-3.13$ -- $-0.642\,\mathrm{rad}$, and clusters 0 and 2 occupy intervals centered closer to zero.
\footnote{A relation equivalent to $\delta_{\rm CP} \simeq \alpha_{31}/2$, up to phase conventions, is obtained in Ref.~\cite{Aizawa:2005yy} within a specific complex neutrino-mass texture in which a single CP phase controls the breaking of $\mu$-$\tau$ permutation symmetry. While this provides an interesting theoretical precedent for a correlation between the Dirac and Majorana CP phases, it remains unclear whether the relation found in our analysis is connected to the same underlying mechanism or arises from a different structure. Clarifying the possible theoretical origin of this correlation, including its relation to previously proposed flavor and CP-symmetry frameworks, would therefore be an important direction for future study.} 
By contrast, the ranges of $\alpha_{21}$ are nearly identical, extending approximately $-0.75$ -- $+0.73\,\mathrm{rad}$ in every cluster.

\medskip

\begin{table}[th]
\centering
\caption{Range of observables in each cluster.}
\label{tab:cluster_ranges}
\resizebox{\textwidth}{!}{%
\begin{tabular}{c c c c c c}
\hline
Cluster
& $m_1\,[\mathrm{meV}]$
& $m_2\,[\mathrm{meV}]$
& $m_3\,[\mathrm{meV}]$
& $\delta_{\mathrm{CP}}\,(\simeq \alpha_{31}/2)\,[\mathrm{rad}] $
& $\alpha_{21}\,[\mathrm{rad}]$ \\
\hline
0 & $0.0144$ -- $6.90$ & $7.99$ -- $10.8$ & $48.5$ -- $52.1$ & $-0.473$ -- $+1.92$ & $-0.692$ -- $+0.733$ \\
1 & $0.0133$ -- $7.36$ & $8.01$ -- $11.4$ & $48.7$ -- $51.7$ & $+1.43$ -- $+3.14$ & $-0.706$ -- $+0.727$ \\
2 & $0.00147$ -- $3.86$ & $7.99$ -- $9.70$ & $48.4$ -- $51.5$ & $-1.14$ -- $+1.17$ & $-0.752$ -- $+0.705$ \\
3 & $0.00559$ -- $5.36$ & $7.92$ -- $10.3$ & $48.7$ -- $51.5$ & $-3.13$ -- $-0.642$ & $-0.703$ -- $+0.719$ \\
\hline
\end{tabular}%
}
\end{table}

\newpage

As a further analysis, for points satisfying $\chi^2 < 45$ within each cluster, we approximate each observable using a quadratic function of $(z_1, z_2)$. 
To prevent the accumulation of errors, $\sum_i m_i$ and $\langle m_{ee} \rangle$ are approximated directly as functions of $(z_1, z_2)$, rather than by combining the approximations of the observables in Eq.~\eqref{eq:def_A}. 
The coefficients of determination are presented in Table \ref{tab:r2_2d}.

\medskip

The following equations present the approximation results for the entries in Table \ref{tab:r2_2d}, where the coefficient of determination is 0.8 or higher. 
Here, $m_1, \langle m_{ee} \rangle, \sum_i m_i$ are expressed in $\meV$, while the CP phases are given in radians. 
Note that $\alpha_{31}$ is not selected when $\delta_{\mathrm{CP}}$ is chosen, since $\delta_{\mathrm{CP}}$ is approximately equal to $\alpha_{31}/2$ for all clusters. 
Additionally, in Cluster 2, the coefficient of determination does not exceed 0.8 for any observables other than $m_1$ and $\delta_{\mathrm{CP}}$, so we select $\langle m_{ee} \rangle$ as having relatively high accuracy.
\begin{description}
    \item[Cluster 0]
    \begin{align}
    \begin{split}
        m_1
        &= 2.72z_1^2 + 1.00z_2^2 - 2.75z_1z_2
        + 12.44z_1 - 6.19z_2 + 13.65, \\
        \delta_{\mathrm{CP}}
        &= -0.74z_1^2 - 0.02z_2^2 + 0.58z_1z_2
        - 4.74z_1 + 0.69z_2 + 85.19, \\
        \langle m_{ee} \rangle
        &= 1.04z_1^2 + 0.73z_2^2 - 1.38z_1z_2
        + 4.58z_1 - 3.54z_2 + 8.79.
    \end{split}
    \label{eq:quad_cluster_0}
    \end{align}

    \item[Cluster 1]
    \begin{align}
    \begin{split}
        m_1
        &= -4.77z_1^2 + 0.85z_2^2 - 6.45z_1z_2
        - 29.22z_1 - 17.27z_2 - 43.82, \\
        \langle m_{ee} \rangle
        &= -5.14z_1^2 + 0.73z_2^2 - 6.66z_1z_2
        - 30.94z_1 - 18.69z_2 - 42.34, \\
        \sum_i m_i
        &= -29.56z_1^2 - 8.59z_2^2 - 44.61z_1z_2
        - 179.76z_1 - 136.94z_2 - 212.26.
    \end{split}
    \label{eq:quad_cluster_1}
    \end{align}

    \item[Cluster 2]
    \begin{align}
    \begin{split}
        m_1
        &= -1.53z_1^2 + 0.08z_2^2 + 0.77z_1z_2
        - 2.36z_1 + 1.57z_2 - 0.02, \\
        \delta_{\mathrm{CP}}
        &= -0.46z_1^2 + 0.07z_2^2 + 0.59z_1z_2
        - 1.58z_1 - 0.47z_2 + 0.34, \\
        \langle m_{ee} \rangle
        &= -1.44z_1^2 - 0.04z_2^2 + 0.80z_1z_2
        - 1.74z_1 + 1.23z_2 + 3.59.
    \end{split}
    \label{eq:quad_cluster_2}
    \end{align}

    \item[Cluster 3]
    \begin{align}
    \begin{split}
        m_1
        &= 3.27z_1^2 + 0.25z_2^2 - 1.33z_1z_2
        - 7.82z_1 + 4.37z_2 - 0.62, \\
        \langle m_{ee} \rangle
        &= 2.88z_1^2 + 0.67z_2^2 - 1.62z_1z_2
        - 5.19z_1 + 0.00z_2 + 3.63, \\
        \sum_i m_i
        &= 15.12z_1^2 + 2.97z_2^2 - 9.68z_1z_2
        - 21.86z_1 + 5.13z_2 + 63.14.
    \end{split}
    \label{eq:quad_cluster_3}
    \end{align}
\end{description}

\medskip

We emphasize that the latent coordinates $z_1$ and $z_2$ themselves do not possess a unique physical meaning. 
Their numerical values, orientations, and even nonlinear parameterizations can vary upon retraining the autoencoder or reparameterizing the latent space. 
Accordingly, the numerical coefficients in Eqs.~\eqref{eq:quad_cluster_0}-\eqref{eq:quad_cluster_3} should not be interpreted as physical invariants. 
Instead, the physical content resides in the image of the latent representation within the observable space. 
For example, a two-dimensional surface parameterized as 
\begin{align}
\left( m_1,\langle m_{ee}\rangle,\delta_{\rm CP} \right) = \mathbf{f}(z_1,z_2)
\end{align}
can equivalently be represented, at least locally, by an implicit relation:
\begin{align}
F \left( m_1,\langle m_{ee}\rangle,\delta_{\rm CP} \right) \simeq 0.
\end{align}
Here, $F$ denotes the relationship defined among physical quantities that specifies the surface constraining their mutual variations.
Under an invertible change of latent coordinates, 
\begin{align}
\mathbf{z} \to \mathbf{z}' = \mathbf{g}(\mathbf{z}),
\end{align}
the parameterization $\mathbf{f}$ changes, while its image in the physical observable space remains invariant. 
Thus, the physically relevant prediction is the correlation surface among the observables, rather than the particular choice of latent coordinates used to parameterize it.
The separation between clusters shown in Fig.~\ref{fig:hdbscan_physical_hist} and the regression accuracy presented in Table \ref{tab:r2_2d} do not necessarily reflect the same characteristics. 
Fig.~\ref{fig:hdbscan_physical_hist} illustrates the extent to which the distributions of each observable differ across clusters. 
In contrast, the coefficient of determination $R^2$ in Table \ref{tab:r2_2d} measures how well variations in a physical quantity can be explained as a function of the latent coordinates $(z_1, z_2)$ within a single cluster. 
Therefore, even if the distribution of an observable overlaps across multiple clusters, high regression accuracy can still be achieved as long as there is a consistent relationship between the distribution and the positions in latent space within each cluster.

\medskip

$m_1$ clearly illustrates this difference. 
In Fig.~\ref{fig:hdbscan_physical_hist}, the distribution of $m_1$ overlaps significantly across clusters, indicating that it is not an observable capable of distinguishing between clusters. 
However, since the $R^2$ value for $m_1$ is large, it can be assumed that $m_1$ varies continuously along a fixed direction in the latent space within each cluster.
Consequently, even if the separation between clusters is not distinct, $m_1$ can be estimated with high precision from the latent coordinates.

\newpage

Conversely, $\delta_{\rm CP}$ and $\alpha_{31}$ play a crucial role in distinguishing between clusters. 
As shown in Fig.~\ref{fig:hdbscan_physical_hist}, each cluster occupies a distinct phase region, with these phases serving as direct indicators that characterize the clusters. 
Furthermore, in Clusters 0 and 2, phase variations within the clusters progress smoothly along the latent coordinates, resulting in high $R^2$ values for both $\delta_{\rm CP}$ and $\alpha_{31}$. 
In contrast, although Clusters 1 and 3 exhibit phase ranges that differ markedly from those of the other clusters, the changes within them are not adequately represented by quadratic functions. 
This suggests that the ability to effectively distinguish clusters and the ability to achieve high-precision regression within a cluster are distinct properties.

\medskip

The three mixing angles exhibit very low $R^2$ values across all clusters. 
Since the mixing angles are inherently constrained to values close to the experimental data as conditions for flow matching and as $\chi^2$ metrics, their distribution ranges within each cluster are narrow, showing no significant differences in Fig.~\ref{fig:hdbscan_physical_hist}. 
In other words, the residual variability in the mixing angles is not strongly associated with changes in the mass scales or CP phase that define the cluster structure. 
Given that a two-dimensional bottleneck cannot retain all input information, it can be inferred that the autoencoder prioritized preserving the large, coordinated variations observed in $m_1$ and the CP phase within the latent space over these small, cluster-independent variations.

\medskip

However, caution is required when interpreting a low $R^2$ value for the mixing angle as directly indicative of a large absolute prediction error. 
Since the coefficient of determination is defined as the ratio of the regression error to the variance of the observed values, physical quantities with a very narrow distribution range may exhibit a low $R^2$ even when the absolute error is small. 
Furthermore, in this analysis, while the masses and CP phases are normalized, the mixing angles are used as input without any transformation.
This difference in numerical scale may have hindered the autoencoder's ability to preserve subtle fluctuations in the mixing angle within the latent space. 
Therefore, the low regression accuracy for the mixing angle should be understood as a result influenced not only by weak physical correlations but also by the effects of two-dimensional compression and preprocessing.

\medskip

\begin{figure}[th]
    \centering
    \includegraphics[width=0.7\linewidth]{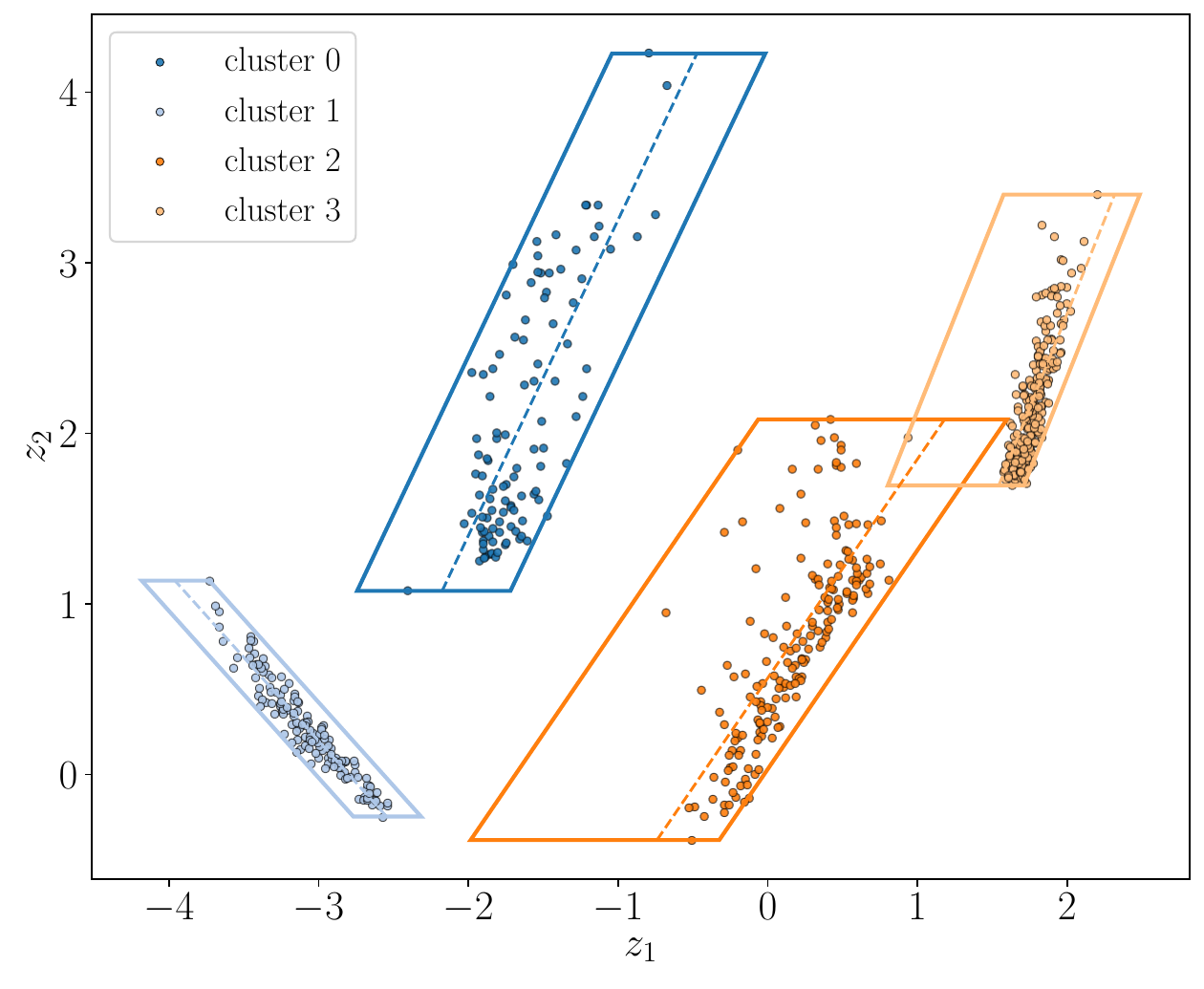}
    \caption{Parallelogram approximations of the four cluster domains in the two-dimensional latent space for samples satisfying $\chi^2<45$. The dashed lines indicate the linear fits used to determine the orientation of each domain.}
    \label{fig:parallelogram_domain}
\end{figure}

\begin{figure}[th]
    \centering
    \includegraphics[width=\linewidth]{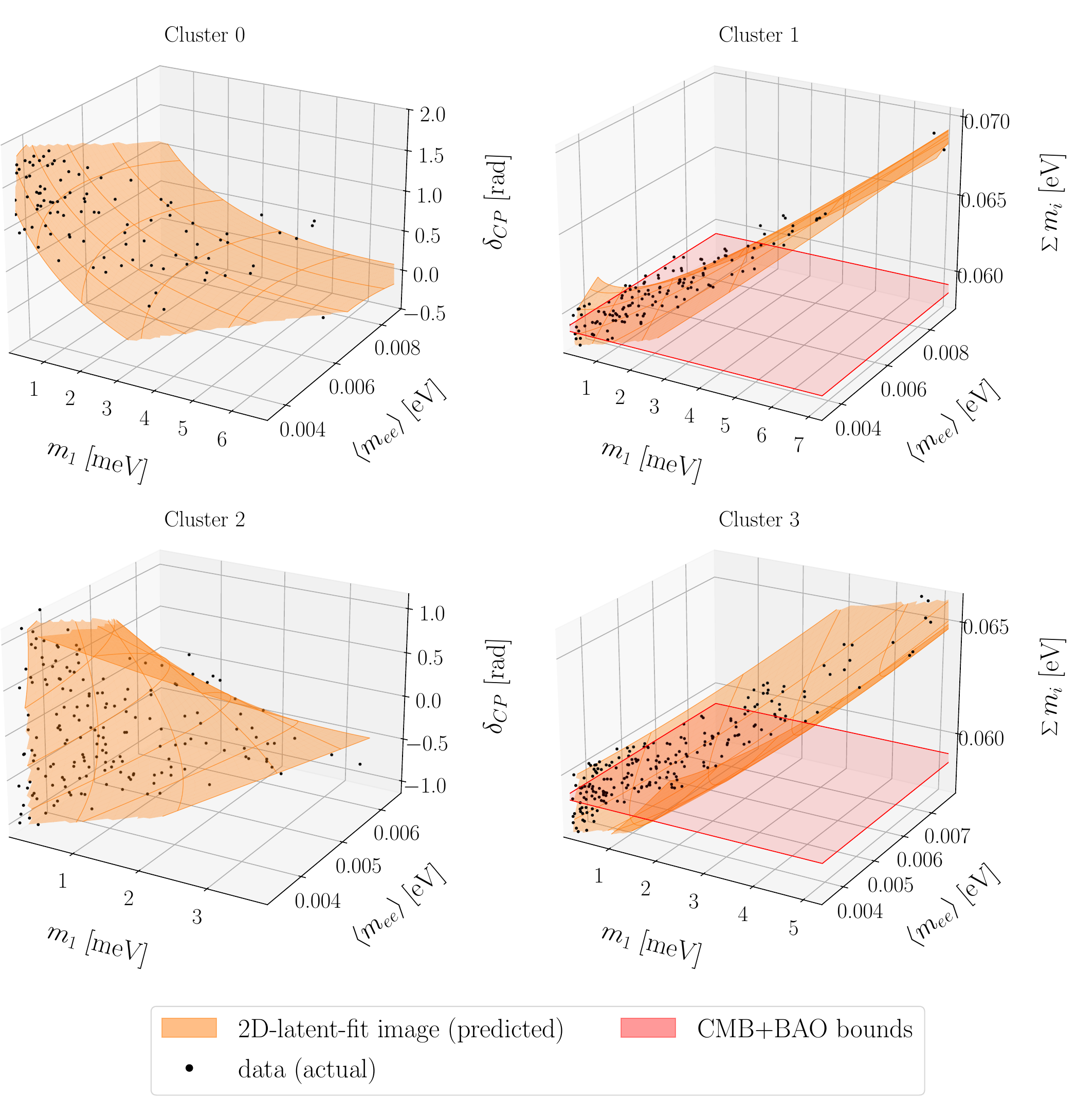}
    \caption{The orange surfaces are obtained by mapping the parallelogram domains shown in Fig.~\ref{fig:parallelogram_domain} through the quadratic regressions of Eqs.~\eqref{eq:quad_cluster_0}–\eqref{eq:quad_cluster_3}, and are truncated to the data ranges shown in Fig.~\ref{fig:hdbscan_physical_hist}. The black points denote the generated samples. The red plane corresponds to the adopted prior to obtain Eq.~\eqref{eq:DESI_BAO+CMB_NO}, which is the constraint for the NO case from CMB+BAO.}
    \label{fig:triples_per_cluster}
\end{figure}

In Fig.~\ref{fig:parallelogram_domain}, the distribution of data satisfying $\chi^2<45$ within each cluster is approximated by a parallelogram in latent space. 
The points inside this parallelogram are then mapped to the observable space using the quadratic functions defined in Eqs.~\eqref{eq:quad_cluster_0}–\eqref{eq:quad_cluster_3}. 
The orange surface in Fig.~\ref{fig:triples_per_cluster} represents the predicted surface obtained through this mapping, while the black dots correspond to the actual generated data. 
Fig.~\ref{fig:triples_per_cluster} therefore provides a representation of the learned correlations directly in terms of physical observables, eliminating the arbitrariness associated with the choice of latent coordinates. 
These figures demonstrate that the three selected observables are not distributed independently but are generally confined to a two-dimensional surface within each cluster. 
The parallelogram does not represent a strict confidence region for the data; rather, it serves as a geometric enclosure of the clusters. 
Therefore, extrapolation effects are significant near the edges, where data density is low, and especially near the vertices of the parallelogram.

\newpage

The key point of Fig.~\ref{fig:triples_per_cluster} is that it clearly demonstrates that the permissible ranges for each physical quantity are localized on two-dimensional correlation surfaces specific to each cluster. 
In other words, this result provides not only predicted ranges for individual observables in future experiments but also consistency conditions for the simultaneous measurement of different observables. 
For example, in Clusters 0 and 2, a correlation surface is formed among $m_1$, $\langle m_{ee}\rangle$, and $\delta_{\rm CP}$. 
Considering a scenario in which precise measurements of $\delta_{\rm CP}$ from neutrino oscillation experiments and constraints on $\langle m_{ee}\rangle$ from future searches for neutrinoless double beta decay are obtained, the remaining $m_1$ is narrowed down to a limited region on the surface. 
Conversely, even if the three independently obtained observables lie individually within their respective allowed ranges, if their combination falls outside this correlation surface, the corresponding cluster structure can be rejected. 
Similarly, in Clusters 1 and 3, a correlation surface emerges among $m_1, \langle m_{ee}\rangle, \sum_i m_i$. The constraints on the sum of masses from neutrinoless double beta decay searches and cosmological observations function not merely as one-dimensional upper limits, but as complementary probes of the same correlation structure from different perspectives.
In this sense, the surface shown in Fig.~\ref{fig:triples_per_cluster} integrates information obtained from oscillation experiments, cosmological observations, and searches for neutrinoless double beta decay into a single multivariate relationship. 
\footnote{Similar consistency relations arise as neutrino mass or mixing sum rules in flavor-symmetry models, where an underlying theoretical framework constrains several observables, as discussed in Refs.~\cite{Barry:2010yk, King:2013psa}. The correlation surfaces obtained here serve an analogous phenomenological role but emerge from the experimentally conditioned ensemble without imposing such sum rules.}
It is expected that future combinations of multiple experiments will directly verify the latent correlation structure itself, as extracted by machine learning in this analysis.

\medskip

\begin{table}[th]
\centering
\caption{Coefficients of determination $R^2$ for quadratic regressions of the physical observables in terms of the latent variables $(z_1,z_2)$, evaluated separately for each cluster. The bold entries indicate three representative results in each cluster.}
\label{tab:r2_2d}
\begin{tabular}{|c|c|c|c|c|}
\hline
 & Cluster 0 & Cluster 1 & Cluster 2 & Cluster 3 \\
\hline
$m_1$                  & \textbf{0.955} & \textbf{0.978} & \textbf{0.793} & \textbf{0.958} \\
$m_2$                  & 0.612 & 0.697 & 0.144 & 0.493 \\
$m_3$                  & 0.335 & 0.697 & 0.476 & 0.707 \\
$s_{12}$               & 0.070 & 0.017 & 0.014 & 0.035 \\
$s_{13}$               & 0.041 & 0.024 & 0.043 & 0.043 \\
$s_{23}$               & 0.038 & 0.030 & 0.066 & 0.042 \\
$\delta_{\mathrm{CP}}$ & \textbf{0.954} & 0.400 & \textbf{0.988} & 0.290 \\
$\alpha_{21}$          & 0.012 & 0.029 & 0.031 & 0.011 \\
$\alpha_{31}$          & 0.955 & 0.405 & 0.989 & 0.288 \\
$\sum_i m_i$           & 0.769 & \textbf{0.889} & 0.375 & \textbf{0.896} \\
$\langle m_{ee}\rangle$& \textbf{0.913} & \textbf{0.925} & \textbf{0.645} & \textbf{0.884} \\
\hline
\end{tabular}
\end{table}

\section{Conclusions}
\label{sec:con}

The origin of flavor structure remains one of the central unresolved problems in particle physics. 
In particular, the relationships among neutrino masses, leptonic mixing parameters, and CP phases are not yet fully understood. 
In this work, we investigated whether machine learning techniques can be combined to explore the parameter space of the Type-I seesaw mechanism and uncover correlations among leptonic observables without imposing a specific flavor symmetry on the neutrino Yukawa matrix.

\medskip

After introducing the Type-I seesaw mechanism in Sec.~\ref{sec:background}, we employed flow-matching posterior estimation to learn the conditional distribution of the free parameters in Sec.~\ref{sec:parameter_optimization_with_flow_matching}, subject to the requirement that the neutrino mass-squared differences and mixing angles reproduce their experimental results. 
From $3\times10^{6}$ generated parameter points, 956,607 samples satisfied $\chi^{2}<2{,}500$, and 640 samples satisfied the more stringent criterion $\chi^{2}<45$. 
Notably, although the original training data contained no samples with $\chi^{2}<45$, the trained flow-matching model generated a substantial number of such high-accuracy solutions. 
This result demonstrates that the generative model does not merely reproduce individual training samples, but learns the nontrivial high-dimensional structure linking the seesaw parameters to the low-energy observables.

\newpage

In Sec.~\ref{sec:correlation_discovery_via_autoencoder}, we subsequently applied an autoencoder to the neutrino masses, mixing angles, and CP phases derived from the generated solutions. 
The learned two-dimensional latent representation revealed four distinct clusters, with the separation among these clusters being particularly pronounced in the CP phases. 
By approximating the data distribution within each cluster and expressing the original observables in terms of the latent variables, we obtained cluster-dependent relations among the leptonic observables. 
These relationships also imply distinguishable cluster-dependent tendencies between $\delta_{\rm CP}$ and both the sum of the neutrino masses and the effective Majorana mass relevant to neutrinoless double-beta decay.

\medskip

These relationships are non-trivial findings uncovered by integrating data generation with feature analysis, derived from the nonlinear correlations between observables. 
Therefore, the combination of flow matching and autoencoders offers a powerful framework not only for efficiently identifying parameter points consistent with experimental observations, but also for extracting hidden structures from the resulting high-dimensional distributions. 
In this sense, our approach provides a data-driven perspective that complements conventional flavor-model studies, which typically start by assuming a specific symmetry or texture.

\medskip

Before concluding our paper, we would like to mention possible directions for future work:
\begin{itemize}
    \item
    We did not consider the renormalization effects, but incorporating their scale evolution would be valuable. 
    Such an analysis could clarify whether the cluster structures and the correlations among neutrino masses, mixing angles, and CP phases identified in the latent space remain stable under scale dependence.
    \item 
    In our analysis, we employed a conventional fully connected autoencoder with a two-dimensional bottleneck. 
    It would be valuable to replace this architecture with a sparse autoencoder\footnote{Ref.~\cite{Dong:2025sae} provides a comprehensive survey of sparse autoencoders.}, which uses an overcomplete latent representation where only a small subset of latent variables is activated for each input. 
    By encouraging the latent features to respond selectively to different structures in the data, such an architecture may yield a more interpretable representation of the physical observables. 
    This approach is expected to produce more distinct clustering among neutrino masses and mixing angles than that shown in Fig.~\ref{fig:hdbscan_physical_hist}. 
    A systematic comparison between dense and sparse autoencoders, including an assessment of the stability of the extracted relations, is left for future work.
\end{itemize}

\acknowledgments

This work was supported in part by JSPS KAKENHI Grant Numbers JP25KJ1927 (S.N.), JP25H01539 (H.O.) and JP26K07087 (H.O.).

\appendix

\section{Formulation of flow matching}
\label{app:flow_matching}

In this appendix, we summarize the formulation of flow matching \cite{Lipman:2023flo} from the viewpoint of flow-based generative models.
In particular, the continuous normalizing flow (CNF) is a continuous-time version of a normalizing flow: it represents the transport of a probability density by the flow generated by an ordinary differential equation (ODE). 
The essential idea of flow matching is to formulate a generative model as a continuous-time flow that transports probability density from a simple reference distribution to the data distribution. 
The velocity field that generates this flow is then learned as a simple regression problem without performing numerical ODE integration during training.
Ref.~\cite{flowsanddiffusions2026} gives a clear review for flow matching.

\medskip

Let the data be $x \in \mathbb{R}^d$, and denote by $p_t(x)$ the probability density at time $t \in [0,1]$. 
Here, $p_0$ is a tractable reference distribution, such as the standard normal distribution, while $p_1$ is a distribution that approximates the data distribution $q$. 
A \textit{flow} $\phi_t(x)$ generated by a \textit{time-dependent vector field} $v_t(x)$ is defined as follows:
\begin{align}
    \frac{d}{dt}\phi_t(x) &= v_t\left(\phi_t(x)\right), \quad
    \phi_0(x) = x.
\end{align}
Then, a \textit{push-forward} operation for probability distribution is defined in the following notation.
\begin{align}
    [\phi_t]_{*} p_0(x) = p_0\left(\phi_{t}^{-1}(x)\right) \det \left[ \frac{\partial \phi_t^{-1}(x)}{\partial x} \right]
\end{align}
This means the probability distribution obtained by transporting samples from $p_0$ through the map $\phi_t$. 
When this relation is satisfied (i.e.~the flow $\phi_t$ pushes forward $p_0$ to $p_t$), the vector field $v_t$ is the velocity field that generates the probability-density path $p_t$. 
Locally, this relation is characterized by the continuity equation
\begin{align}
    \frac{\partial p_t(x)}{\partial t} + \operatorname{div} \left[p_t(x) v_t(x)\right] = 0.
\end{align}
This equation has the same structure as the conservation law that appears in fluid dynamics. 
It states that the probability density is transported conservatively along the vector field.

\medskip

Suppose that the desired probability-density path $p_t$ and the corresponding velocity field $u_t(x)$ are given. 
You can train a neural network representing a vector field $v_t(x;\theta)$ by minimizing the following loss function:
\begin{align}
    L_{\mathrm{FM}}(\theta) =
    \mathbb{E}_{t \sim \mathcal{U}[0,1],\, x \sim p_t}
    \left[ \left| v_t(x;\theta) - u_t(x) \right|^2 \right].
\end{align}
If this loss becomes sufficiently small, the learned ODE reproduces the transport from $p_0$ to $p_1$. 
Therefore, a sample of the data distribution is obtained by drawing an initial point from $p_0$ and integrating the ODE to $t=1$. 
However, the density of the data distribution $q$ is unknown in practice, so $p_t$ and $u_t$ cannot be calculated directly. 
To avoid this difficulty, conditional probability paths and conditional velocity fields are introduced.

\medskip

For a data point $x_1 \sim q$, consider a conditional probability path $p_t(x|x_1)$. 
This is a family of distributions that coincides with the reference distribution $p_0$ at $t=0$ and concentrates near $x_1$ at $t=1$. 
The following choice is typical one.
\begin{align}
\begin{split}
    p_0(x|x_1) &= p_0(x), \\
    p_1(x|x_1) &\simeq \mathcal{N}(x|x_1,\sigma_{\min}^2 I),
\end{split}
\end{align}
where $\mathcal{N}(x|x_1,\sigma_{\min}^2 I)$ is a Gaussian distribution with the mean $x_1$ and the variance $\sigma_{\min}^2 I$.
By integrating this conditional path over the data distribution $q(x_1)$, one obtains the marginal probability path as follows:
\begin{align}
    p_t(x) = \int dx_1\,p_t(x|x_1) q(x_1).
\end{align}
At $t=1$, this becomes a mixture of distributions concentrated near each data point. 
Hence, $p_1$ approximates $q$ if $\sigma_{\min}$ is sufficiently small.

\medskip

If the conditional velocity field $u_t(x|x_1)$ that generates the conditional path $p_t(x|x_1)$ is known, then the velocity field $u_t(x)$ can formally be written as follows:
\begin{align}
    u_t(x) = \int dx_1\,u_t(x|x_1) \frac{p_t(x|x_1)q(x_1)}{p_t(x)}.
\end{align}
This expression implies that the velocity field can be interpreted as the posterior average of the conditional velocity fields pointing toward the various data points.
Thus, $u_t(x)$ generates the marginal probability path $p_t(x)$ by satisfying the continuity equation. However, the expression above contains $p_t(x)$ and therefore cannot be evaluated directly in practice.

\medskip

It is a significant feature that flow matching does not need to explicitly compute this intractable velocity field $u_t(x)$. 
Instead, regression onto the conditional velocity field $u_t(x|x_1)$ gives the same training gradient. 
In other words, when you consider the conditional flow matching loss
\begin{align}
    L_{\mathrm{CFM}}(\theta) =
    \mathbb{E}_{t \sim \mathcal{U}[0,1],\,x_1 \sim q,\, x \sim p_t(\cdot|x_1)}
    \left[ \left| v_t(x;\theta) - u_t(x|x_1) \right|^2 \right],
\end{align}
$L_{\mathrm{CFM}}$ and $L_{\mathrm{FM}}$ are equivalent up to a constant independent of $\theta$ and the following result is derived.
\begin{align}
    \nabla_\theta L_{\mathrm{CFM}}(\theta) = \nabla_\theta L_{\mathrm{FM}}(\theta).
\end{align}
Therefore, during training, you only need to sample a data point $x_1$, a time $t$, and a point $x$ on the conditional path, and then perform squared-error regression using the known conditional velocity field as the target. 
There is no need to solve an ODE sequentially to train the model. 
Hence, the training procedure is simulation-free.

\medskip

In summary, this mechanism can be understood as follows. 
The conditional path $p_t(x|x_1)$ is a virtual transport channel whose endpoint is the data point $x_1$, and each channel is assigned a local velocity $u_t(x|x_1)$. 
In the actual generative process, the individual endpoint $x_1$ is not fixed. 
Instead, all channels are superimposed according to the data distribution $q$, and then, the resulting effective velocity field is $u_t(x)$. 
Conditional flow matches only one component of this superposition in each minibatch and uses its local velocity as the training target. 
After taking the expectation, this is equivalent to learning the effective velocity field produced by the full superposition. 
The teacher velocity for an individual sample is not itself equal to the velocity field at a fixed $x$. 
Nevertheless, the gradient of the squared loss averaged over $(t,x_1,x)$ agrees with the gradient of the flow matching loss for the marginal velocity field.

\medskip

The conditional probability path most commonly used in implementations is a Gaussian path. 
Using a time-dependent mean $\mu_t(x_1)$ and standard deviation $\sigma_t(x_1)$, it is defined as follows:
\begin{align}
    p_t(x|x_1) =
    \mathcal{N}\left(x \middle| \mu_t(x_1), \sigma_t(x_1)^2 I\right).
\end{align}
When we impose the boundary conditions
\begin{align}
\begin{split}
    \mu_0(x_1) &= 0, \quad
    \sigma_0(x_1) = 1, \\
    \mu_1(x_1) &= x_1, \quad
    \sigma_1(x_1) = \sigma_{\min},
\end{split}
\end{align}
the conditional path becomes the standard normal distribution at $t=0$ and a distribution concentrated near $x_1$ at $t=1$. 
The following conditional flow is a simple form satisfying the above path.
\begin{align}
    \psi_t(x_0) = \sigma_t(x_1)x_0 + \mu_t(x_1), \label{eq:affine_map}
\end{align}
while the corresponding conditional velocity field is determined as follows:
\begin{align}
    u_t(x|x_1) =
    \frac{\sigma_t'(x_1)}{\sigma_t(x_1)} \left[x-\mu_t(x_1)\right]
    + \mu_t'(x_1).
\end{align}
Here, $x_0$ is sampled as $x_0 \sim \mathcal{N}\left(0, I\right)$ and the prime denotes differentiation with respect to $t$. 
This formula is obtained simply by differentiating Eq.~\eqref{eq:affine_map} with respect to time and rewriting the result as a function of the current position $x=\psi_t(x_0)$. Thus, for Gaussian paths, the conditional velocity field can be used directly as the regression target.

\medskip

A particularly important choice is to make both the mean $\mu_t$ and the standard deviation $\sigma_t$ linear in time:
\begin{align}
    \mu_t(x_1) &= t x_1, \quad
    \sigma_t(x_1) = 1-(1-\sigma_{\min})t.
\end{align}
In this case, the conditional flow and the conditional velocity field are derived as follows:
\begin{align}
\begin{split}
    \psi_t(x_0) &= \left[1-(1-\sigma_{\min})t\right]x_0 + t x_1, \\
    u_t(x|x_1) &= \frac{x_1-(1-\sigma_{\min})x} {1-(1-\sigma_{\min})t}.
\end{split}
\end{align}
Substituting $x=\psi_t(x_0)$, the teacher velocity used during training becomes
\begin{align}
    \frac{d}{dt}\psi_t(x_0) = x_1-(1-\sigma_{\min})x_0.
\end{align}
Therefore, the conditional flow matching loss in this case is written as the following expression.
\begin{align}
    L_{\mathrm{CFM}}(\theta) =
    \mathbb{E}_{t,x_1,x_0}
    \left[
    || v_t\left(
    x_t;\theta
    \right) -
    \left\{x_1-(1-\sigma_{\min})x_0\right\}
    ||^2
    \right].
\end{align}
In the approximation $\sigma_{\min}\to 0$, this reduces to regression onto a straight-line interpolation with constant velocity:
\begin{align}
    x_t = (1-t)x_0 + t x_1, \quad
    u_t = x_1 - x_0.
\end{align}
This linear path is the conditional displacement interpolation corresponding to the optimal transport between two Gaussian distributions. 
In other words, each noise point $x_0$ moves linearly toward the data point $x_1$ at constant velocity. 
Note that the term corresponding to the optimal transport is justified only for the conditional flow with $x_1$ fixed. 
After marginalizing over the data distribution, the resulting effective flow is not necessarily the global optimal transport map between $p_0$ and $q$.

\medskip

The generative process of flow matching appears as ODE integration only after training. 
Namely, using the learned velocity field $v_t(x;\theta)$, the following equation is numerically solved from $t=0$ to $t=1$,
\begin{align}
    X_0 \sim p_0, \quad
    \frac{d}{dt}X_t = v_t(X_t;\theta),
\end{align}
and $X_1$ is regarded as a generated sample.
Although the ODE is not solved during training, an ODE solver is used during inference process.
This separation avoids the expensive trajectory integration required when training CNFs by maximum likelihood, while still allowing the trained model to be treated as a deterministic continuous-time generative model.

\medskip

In summary, flow matching is a method for learning an unknown velocity field not by observing it directly, but by introducing analytically controllable auxiliary problems called conditional paths and sampling their local velocities. 
This is close in spirit to constructing the effective equation of motion of a complicated many-body system as an ensemble average over simpler conditional motions. 
Each conditional problem is solvable, and their mixture gives the desired effective dynamics.

\medskip

From this viewpoint, the design of flow matching is concentrated in the choice of probability path.
The intermediate distributions $p_t$ connecting $p_0$ to $q$ are not unique. 
The chosen path affects the smoothness of the velocity field, the stiffness of the ODE, the ease of numerical integration, and the simplicity of the learning target. 
Here, stiffness means numerical difficulty caused by widely separated time scales or rapidly varying vector fields. 
Linear paths based on optimal transport are useful because the conditional trajectories are straight lines and the time dependence of the velocity field is relatively simple. 
Ref.~\cite{Lipman:2023flo} reports that flow matching with optimal-transport-type paths is advantageous for image generation in terms of both training efficiency and sampling efficiency.

\medskip

In conclusion, flow matching consists of the following three steps. 
First, one designs a conditional probability path $p_t(x|x_1)$ from the reference distribution toward the neighborhood of each data point. 
Second, one analytically derives the velocity field $u_t(x|x_1)$ that generates this conditional path. 
Third, one samples data points and regresses the neural-network velocity field $v_t(x;\theta)$ onto $u_t(x|x_1)$. 
Hence, the trained model functions as a CNF that generates probability-mass transport from $p_0$ to $q$.

\medskip

The FMPE implementation in \textit{sbi}, which is specialized for conditional posterior estimation, rigorously differs from the generic formulation of flow matching. 
Specifically, it learns a vector field on the parameter space conditioned on simulated observations, rather than an unconditional vector field on the data space. 
The implementation uses a linear interpolation between posterior samples and a Gaussian reference distribution, together with implementation-specific normalization and numerical stabilization. 
These details affect the posterior sampler used in this work but do not change the role of FMPE in our pipeline, namely to provide samples from an approximate posterior distribution.

\bibliography{references}{}
\bibliographystyle{JHEP} 

\end{document}